\documentclass[twocolumn, reprint, nofootinbib]{revtex4-1}
\usepackage{amsmath, amsfonts, amssymb, array}
\usepackage{natbib}
\usepackage{mathrsfs}

\usepackage{slashed}
\usepackage[hidelinks]{hyperref}

\hypersetup{
  colorlinks   = true, 
  urlcolor     = blue, 
  linkcolor    = blue, 
  citecolor   = blue 
}

\begin{document}
\title{A 24+24 real scalar multiplet in four dimensional N=2 conformal supergravity}

\author{Subramanya Hegde$^{1}$, Ivano Lodato$^{2}$ and Bindusar Sahoo$^{1}$}

\affiliation{$^{1}$Indian Institute of Science Education and Research,
Thiruvananthapuram, Vithura, Kerala, 695551, India\\
$^{2}$ Department of Physics and Center for Field Theory and Particle Physics,
Fudan University,
220 Handan Road, 200433 Shanghai, China } 

\begin{abstract}
Starting from the 48+48 component multiplet of supercurrents for a rigid N=2 tensor multiplet in four spacetime dimensions, we obtain the transformation of the linearized supergravity multiplet which couples to this supercurrent multiplet. At the linearized level, this 48+48 component supergravity multiplet decouples into the 24+24 component linearized standard Weyl multiplet and a 24+24 component irreducible matter multiplet containing a real scalar field. By a consistent application of the supersymmetry algebra with field dependent structure constants appropriate to N=2 conformal supergravity, we find the full transformation law for this multiplet in a conformal supergravity background. By performing a suitable field redefinition, we find that the multiplet is a generalization of the flat space multiplet obtained by Howe \textit{et al} in Nucl. Phys. B214 (1983) 519-531, to a conformal supergravity background. We also present a set of constraints which can be consistently imposed on this multiplet to obtain a restricted minimal 8+8 off-shell matter multiplet. We also show as an example the precise embedding of the tensor multiplet inside this multiplet.
\end{abstract}

\allowdisplaybreaks
\maketitle
\section{Introduction}\label{intro}
The use of conformal symmetries plays a crucial role in the construction of theories of supergravity. While the physical Poincar\'e theory can be obtained via a simple gauge-fixing procedure, the higher degree of symmetry in the conformal theories allows for a rigorous re-arrangement of the off-shell degrees of freedom within multiplets shorter than the one that contains Poincar\'e supergravity degrees of freedom. Furthermore, techniques generally referred as ``multiplet calculus'' have been developed over the years to simplify the often massive task of constructing invariant couplings of supergravity, especially higher derivative ones.

Off-shell conformal supergravities exist for dimensions $D\leq 6$. The higher dimensional theories are typically related to the lower dimensional theory via dimensional reduction. For a connection between the $N=(1,0)$ chiral theory in 6 dimensions as well as $N$=$2$ theory in four dimensions with the $N=1$ theory in five dimensions\footnote{We refer to the minimal supergravity in five dimensions formulated in terms of eight supercharges as N=1. However sometimes, for example in \cite{Bergshoeff:2001hc, Fujita:2001kv, Hanaki:2006pj}, it is referred to as N=2.} via dimensional reduction (uplift) see \cite{Kugo:2000hn} and \cite{Banerjee:2011ts} respectively. Hence, one can expect that given a result in higher dimensional theories, a similar result would exist in the lower dimensional theory from dimensional reduction arguments. A relevant example of such connection is the existence of the two versions of Weyl multiplets: the standard and dilaton Weyl, which were constructed in six dimensions \cite{Bergshoeff:1985mz} and later on in five dimensions \cite{Bergshoeff:2001hc,Fujita:2001kv}, also exists in the four-dimensional case \cite{Muller:1986ts,Muller:1986ku,Siegel:1995px,Butter:2017pbp}. While the precise reduction procedure and identification of off-shell degrees of freedom is being currently worked out, it is easy to realize that the reduction of the 32+32 off-shell five-dimensional dilaton Weyl multiplet would give the 24+24 four-dimensional dilaton-Weyl multiplet plus a short 8+8 vector matter multiplet in four dimensions.

It is worth pointing out that the dilaton-Weyl multiplet could be obtained in two different yet equivalent ways in five dimensions. The first procedure involves analyzing the coupling of a standard Weyl multiplet with an on-shell vector multiplet \cite{Bergshoeff:2001hc}. The vanishing of the vector multiplet equations of motion furnishes a set of constraints, which can be used to fix the auxiliary fields of the standard Weyl multiplet in terms of the auxiliary fields of the dilaton Weyl multiplet. The same procedure is used in \cite{Butter:2017pbp} to obtain the $N=2$ dilaton-Weyl multiplet in four dimensions. 

The second method in five dimensions was based on the construction of the current multiplet for a non-conformal vector multiplet. The standard action for a non-conformal vector multiplet in five dimensions would not be scale invariant and hence the trace of the energy-momentum tensor $\sigma=\theta^\mu_\mu$ would be a non-trivial component of the current multiplet. When coupling the current multiplet to linearized gravity (Weyl) multiplet, the scalar current $\sigma$ will couple to a scalar field of dimension 1: the dilaton. From the linearized gravity multiplet it is in principle a simple, yet computationally challenging, task to derive the full non-linear transformations of the dilaton version of the Weyl multiplet. It is worth emphasizing that, in this method, the non-conformal nature of the matter multiplet chosen was crucial to obtain the Weyl multiplet with a dilaton.

One na\"ively expects the same method to produce the dilaton Weyl multiplet in four dimensions, the only caveat being that the vector multiplet in four dimensions is conformally invariant. Hence a different off-shell non-conformal multiplet needs to be chosen for the construction of the current multiplet. Such non-conformal current multiplets have been discussed in \cite{Stelle:1981gi}. We use the current multiplet associated with a rigid tensor multiplet in four spacetime dimensions and show that the na\"ive expectation of getting a dilaton Weyl multiplet from this non-conformal current multiplet turns out to be incorrect.

Specifically, we find that the linearized multiplet of gravity that couples to the current multiplet of the rigid tensor multiplet has 48+48 components, which can be arranged into the standard Weyl multiplet and an irreducible and unconstrained off-shell 24+24 matter multiplet containing a real scalar field. We will henceforth refer to it as the ``real scalar multiplet''. After a suitable field redefinition, we show that our result is a generalization of the flat space results of \cite{Howe:1982tm} to a conformal supergravity background. We also find that this multiplet can be consistently restricted to an 8+8 component multiplet and show the precise identifications encoding the embedding of the tensor multiplet inside this multiplet.
 
The paper is organized as follows. In section \ref{Weyl}, we will describe the features of four-dimensional conformal supergravity relevant to our paper (see \cite{Bergshoeff:1980is} for all the details and \cite{Mohaupt:2000mj} for an extensive review on the subject). In sections \ref{tensor} and \ref{lin_sugra} we will present the details of the construction of the linearized real scalar multiplet starting from the current multiplet of a rigid tensor multiplet. In section \ref{result}, we will present the real scalar multiplet along with its complete supersymmetry transformations in a conformal supergravity background. In section-\ref{field-redef}, we will introduce field redefinitions which will simplify the supersymmetry transformations presented in section-\ref{result}. This is one of the main results of our paper which generalizes the flat space results of \cite{Howe:1982tm} to a conformal supergravity background. In section \ref{con_mul}, we will find a set of consistent constraints that restricts the multiplet to 8+8 components. Finally in section \ref{tensor_new}, we will find the embedding of the 8+8 tensor multiplet within the 24+24 real scalar multiplet that satisfies the constraints obtained in the previous section. The results obtained in section-\ref{con_mul} and \ref{tensor_new} are other important results of our paper which has potential applications that we will discuss in the concluding section.

\section{N=2 conformal supergravity in four spacetime dimensions}\label{Weyl}

Conformal supergravity with N-extended supersymmetry is a gauge theory based on SU$(2,2|N)$ superconformal algebra, the supersymmetric generalization of the conformal algebra. It is invariant under two different types of supersymmetry generators, the Q-supersymmetry and the special S-supersymmetry. The multiplet of fields that contains the gauge fields of the superconformal algebra is known as Weyl multiplet. For $N=2$, the Weyl multiplet has 24+24 (bosonic+fermionic) off-shell degrees of freedom. In order to describe the multiplet, we fix our notations as follows. We denote the spacetime indices by Greek letters $\mu,\nu\cdots$, the local Lorentz indices by Latin letters $a,b,\cdots$, SU(2) indices by $i,j\cdots$. As we mentioned before, two different versions of the Weyl multiplet exists in four dimensions: here we will focus on the standard formulation. The field content of the standard Weyl multiplet contains the gauge fields of the SU$(2,2|2)$ algebra: $e_{\mu}{}^{a}$ (vielbein), an SU(2) doublet of Majorana spinors whose positive and negative chiral projections are denoted by $\psi_{\mu}{}^{i}$ and $\psi_{\mu}{}_{i}$ respectively (gravitinos), an SU(2) triplet of gauge fields $V_{\mu}{}^{i}{}_{j}$ corresponding to SU$(2)$ R-symmetry, gauge field $A_{\mu}$ corresponding to U(1) R-symmetry, gauge field $b_{\mu}$ corresponding to the dilatation symmetry. Apart from the above-mentioned gauge fields, the Weyl multiplet also contains several auxiliary fields that are required to balance bosonic and fermionic degrees of freedom. They include an SU(2) doublet of Majorana spinors whose positive and negative chiral projections are denoted by $\chi^{i}$ and $\chi_{i}$ respectively, a real tensor $T_{ab}$ anti-symmetric in its Lorentz indices and a real scalar field $D$. The above-mentioned field content of the Weyl multiplet can be tabulated in Table-\ref{Table-Weyl} along with their chiral and Weyl weights. 
\begin{table}
\caption{}\label{Table-Weyl}
\begin{center}
  \begin{tabular}{ | p{1cm}|p{2cm}|p{1cm}|p{1cm}|p{2cm}| }
\hline
    Field & SU(2) Irreps & Weyl weight (w) & Chiral weight (c) & Chirality fermions \\ \hline
    $e_{\mu}{}^{a}$ & $\bf{1}$ & -1 & 0 & -- \\ \hline
    $V_{\mu}{}^{i}{}_{j}$ & $\bf{3}$ & 0 & 0 & -- \\ \hline
    $A_{\mu}$ & $\bf{1}$ & 0 & 0 & -- \\ \hline
    $b_{\mu}$ & $\bf{1}$ & 0 & 0 & -- \\ \hline
    $T^{-}_{ab}$ & $\bf{1}$ & 1 & -1 & -- \\ \hline
    $D$ & $\bf{1}$ & 2 & 0 & -- \\ \hline
    $\psi_{\mu}{}^{i}$ & $\bf{2}$ & -1/2 & -1/2 & +1 \\ \hline
    $\chi^{i}$ & $\bf{2}$ & 3/2 & -1/2 & +1 \\ \hline
  \end{tabular}
\end{center}
\end{table}
Here $T^{\mp}_{ab}$ refers to the (anti)self-dual components of $T_{ab}$. Complex conjugation is used to raise/lower SU(2) indices and change the duality properties of a tensor. It hence flips the chiral weight as well as chirality (for fermions) but not the Weyl weight.
The gauge fields $\omega_{\mu}{}^{ab}$ corresponding to local Lorentz transformations, $f_{\mu}{}^{a}$ corresponding to special conformal transformation and $\phi_{\mu}{}_{i}$ corresponding to S-supersymmetry are dependent and are determined by the following set of conventional constraints:
\begin{align}\label{constraints}
R(P)_{\mu\nu}{}^{a}&=0\;,
\nonumber \\
\gamma^{\mu}\left(R(Q)_{\mu\nu}{}^{i}+\frac{1}{2}\gamma_{\mu\nu}\chi^{i}\right)&=0\;, \nonumber \\
e^{\nu}{}_{b}R(M)_{\mu\nu}{}_{a}{}^{b}-i\tilde{R}(A)_{\mu a}+\frac{1}{4}T^{+}{}_{ab}T^{-}{}_{\mu}{}^{b}-\frac{3}{2}De_{\mu a}&=0\;.
\end{align}
The super-covariant curvature $R(P)$, $R(Q)$ and $R(M)$ appearing above are associated with local translation, Q-supersymmetry and local Lorentz transformations respectively and $\tilde{R}(A)$ is the dual of the curvature associated with the U(1) R-symmetry. The supersymmetry commutators, in particular $[\delta_Q, \delta_Q]$, gets modified by field dependent structure constants as mentioned below:
\begin{align}\label{algebra}
\left[\delta_Q(\epsilon_1),\delta_Q(\epsilon_2)\right]&=\delta^{(cov)}(\xi)+\delta_{M}(\varepsilon)+\delta_{K}(\Lambda_K)+\delta_S(\eta)\nonumber \\
&\quad +\delta_{gauge}\;,
\nonumber \\
&\nonumber \\
\left[\delta_S(\eta),\delta_Q(\epsilon)\right]&=\delta_{M}(\bar{\eta}_{i}\gamma^{ab}\epsilon^{i}+\text{h.c})+\delta_{D}(\bar{\eta}^{i}\epsilon_{i}+\text{h.c})\nonumber \\
& \quad +\delta_A(i\bar{\eta}_{i}\epsilon^{i}+\text{h.c})
\nonumber \\
&\quad +\delta_{V}(-2\bar{\eta}^{i}\epsilon_{j}-(\text{h.c ; traceless}))\;, \nonumber \\
\left[\delta_S(\eta_1),\delta_S(\eta_2)\right]&=\delta_{K}(\bar{\eta}_{2i}\gamma^a \eta_{1}^{i}+\text{h.c})\;.
\end{align}
The infintesimal transformations, $\delta_Q$, $\delta_S$, $\delta_M$, $\delta_K$, $\delta_D$, $\delta_A$ and $\delta_V$, correspond to Q-supersymmetry, S-supersymmetry, local Lorentz, special conformal, dilatation, U(1) and SU(2) R-symmetry respectively. The infinintesimal covariant transformation, $\delta^{(cov)}$ is defined as:
\begin{align}\label{deltacovdef}
\delta^{(cov)}(\xi)&= \delta_{gct}(\xi)+\sum_{T}\delta^{T}(-\xi^{\mu}h_{\mu}(T))\;,
\end{align}
where $\delta_{gct}$ is general coordinate transformation. The sum is over all superconformal transformation (and any other gauge transformation present in the multiplet) except local translation and the field $h_{\mu}(T)$ is the corresponding gauge field, except for Q and S-supersymmetry, where $h_{\mu}(T)$ is $\frac{1}{2}$ times the corresponding gauge fields and for SU(2) R-symmetry where $h_{\mu}(T)$ is $-\frac{1}{2}$ times the corresponding gauge field. Finally the effect of any other gauge transformation (for example in vector multiplet) is encoded in $\delta_{gauge}$ that appears on the RHS of $[\delta_Q,\delta_Q]$ algebra in (\ref{algebra}). The parameters that appear above in (\ref{algebra}) read:
\begin{align}\label{parameters_algebra}
\xi^{\mu}&=2\bar{\epsilon}_{2}^{i}\gamma^{\mu}\epsilon_{1i}+\text{h.c.}\;,
\nonumber \\
\varepsilon_{ab}&=\varepsilon^{ij}\bar{\epsilon}_{1i}\epsilon_{2j}T^{-}_{ab}+\text{h.c.}\;,\nonumber \\
\Lambda_{K}^{a}&=\varepsilon^{ij}\bar{\epsilon}_{1i}\epsilon_{2j}D_{b}T^{-}{}^{ba}-\frac{3}{2}\bar{\epsilon}_{2}^{i}\gamma^{a}\epsilon_{1i}D+\text{h.c.}\;,
\nonumber \\
\eta^{i}&=6\bar{\epsilon}_{1}^{[i}\epsilon_{2}^{j]}\chi_{j}\;.
\end{align}
 Apart from the Weyl multiplet(s), there exists a large variety of $N=2$ superconformal matter multiplets in four dimensions \cite{Fayet:1975yi,Grimm:1977xp,deWit:1979xpv,deWit:1980lyi,Sohnius:1980it,Brandt:2000uw,deWit:2006gn}. These multiplets all contain 8+8 off-shell degrees of freedom, with the exception of the hypermultiplet which is an on-shell multiplet\footnote{The existence of off-shell hypermultiplet is subtle. One can formulate it in the presence of an off-shell central charge. Alternatively one can also formulate it in harmonic superspace \cite{Galperin:1984av}, without the need for a central charge. The latter formulation requires infinite degrees of freedom.}.
  
In section-\ref{result}, we will present a large $N=2$ off-shell matter multiplet, which contains instead 24+24 degrees of freedom and forms an irreducible representation of the SU$(2,2|2)$ superconformal algebra. 

 \section{Tensor multiplet and the multiplet of supercurrents}\label{tensor}
 The current multiplet for a rigid N=2 tensor multiplet in four dimensions has been given in \cite{Stelle:1981gi} as a reducible multiplet containing two parts: a conformal supercurrent $V$ and a trace supermultiplet $L_{ijkl}$ satisfying the superspace constraints\footnote{See also \cite{Butter:2010sc} for a complete classification of all N=2 supercurrent multiplets in four spacetime dimensions.}.
 \begin{align}\label{supercurrent_new}
 D^{ij}V=-\frac{4}{5}D_{kl}L^{ijkl}
 \end{align}
The supersymmetry transformation of the components of this current multiplet can be read off from the above superspace constraints. The components of this current multiplet and its supersymmetry transformations are central to our construction of the real scalar multiplet in four-dimensional N=2 conformal supergravity. Hence, for clarity of exposition, we will discuss in this section the details of this current multiplet construction, in component notation.
  
An off-shell rigid N=2 tensor multiplet in four spacetime dimensions \cite{deWit:2006gn} encompasses 8+8 degrees of freedom, organized in an SU(2) triplet of pseudo-real scalars $L_{ij}$ satisfying the reality constraint $(L_{ij})^{*}\equiv L^{ij}=\varepsilon^{ik}\varepsilon^{jl}L_{kl}$; an SU(2) doublet of Majorana spinors, whose positive components are denoted by $\phi^{i}$; a real tensor gauge field $E_{\mu\nu}$ with vector gauge transformation $\delta E_{\mu\nu} =2\partial_{[\mu}\Lambda_{\nu]}$. To close the algebra off-shell, an auxiliary complex scalar field $G$ is required. When $G$ is set to zero through its algebraic equations of motion, the supersymmetry transformations of the tensor multiplet components are given by :
\begin{align}\label{susy_tensor}
\delta E_{\mu\nu}&=i\bar{\epsilon}^{i}\gamma_{\mu\nu}\phi^{j}\varepsilon_{ij}+\text{h.c.}\;, \nonumber \\
\delta\phi^{i}&=\slashed{\partial}L^{ij}\epsilon_{j}+\varepsilon^{ij}\slashed{H}\epsilon_{j}\;,\nonumber \\
\delta L^{ij}&=2\bar{\epsilon}^{(i}\phi^{j)}+2\varepsilon^{ik}\varepsilon^{j\ell}\bar{\epsilon}_{(k}\phi_{\ell)}\;.
\end{align}
The action that is invariant under the above susy transformation reads:
\begin{align}\label{action_tensor}
S=\int d^4 x\left[H_{\mu}H^{\mu}-\bar{\phi}^{i}\overleftrightarrow{\slashed{\partial}}\phi_{i}-\frac{1}{2}\partial_{\mu}L^{ij}\partial^{\mu}L^{ij}\right]\;,
\end{align}
where we defined
\begin{align}\label{def}
H^{\mu}=\frac{i}{6}\varepsilon^{\mu\nu\rho\sigma}H_{\nu\rho\sigma}\;.
\end{align}
The three form $H_{\mu\nu\rho}$ is the gauge-invariant field strength of the tensor gauge field $E_{\mu\nu}$ defined as $H_{\mu\nu\rho}\equiv 3\,\partial_{[\mu}E_{\nu\rho]}$. The equations of motion for the fields are given by:
\begin{align}\label{eom_tensor}
\slashed{\partial}\phi^{i}=0\;, \quad \partial^{2}L_{ij}=0\;, \quad \partial_{[\mu}H_{\nu]}=0\;.
\end{align}
The tensor gauge field also satisfies the Bianchi identity $\partial^{\mu}H_{\mu}=0$. Apart from the above mentioned supersymmetry transformation (\ref{susy_tensor}), the action (\ref{action_tensor}) is also invariant under an SU$(2)\times$ U$(1)$ R-symmetry.\\
 The current multiplet associated to the tensor multiplet action contains the following components: the energy-momentum tensor $\theta_{\mu\nu}$ arising from translation invariance, a supersymmetry current $J_{\mu i}$ arising from the supersymmetry invariance, an SU(2) current $v_{\mu}{}^{i}{}_{j}$ and a U(1) current $a_{\mu}$ arising from the SU(2)$\times$U(1) R-symmetry. Since the multiplet is non-conformal, the trace $\sigma$ of the energy-momentum tensor and the gamma-trace $\lambda_{i}$ of the supersymmetry current will also be a part of the current multiplet. The above-mentioned components of the current multiplet in terms of the tensor multiplet fields are given as below:
\begin{align}\label{components_current}
\theta_{\mu\nu}&=H_{\mu}H_{\nu}-\frac{1}{2}\eta_{\mu\nu}H^{2}+\frac{1}{2}\bar{\phi}^{i}\gamma_{(\mu}\overleftrightarrow{\partial}_{\nu)}\phi_{i}+\frac{1}{2}\partial_{\mu}L^{ij}\partial_{\nu}L_{ij} \nonumber \\
&\quad -\frac{1}{4}\eta_{\mu\nu}\partial_{\rho}L^{ij}\partial^{\rho}L_{ij}\;,
\nonumber \\
\sigma&\equiv\theta^{\mu}_{\mu}=-H^2-\frac{1}{2}\partial^{\mu}L^{ij}\partial_{\mu}L_{ij}\;,
 \nonumber \\
v_{\mu}{}^{i}{}_{j}&=\frac{1}{8}L^{ik}\overleftrightarrow{\partial_{\mu}}L_{jk}-\frac{1}{4}\left(\bar{\phi}^{i}\gamma_{\mu}\phi_{j}-\frac{1}{2}\delta^{i}_{j}\bar{\phi}^{k}\gamma_{\mu}\phi_{k}\right)\;,
\nonumber \\
a_{\mu}&=-\frac{i}{8}\bar{\phi}^{l}\gamma_{\mu}\phi_{l}\;,
\nonumber \\
J_{\mu}{}_{i}&=\frac{1}{2}\gamma_{\mu}\slashed{H}\phi^{j}\varepsilon_{ij}-H_{\mu}\phi^{j}\varepsilon_{ij}+\phi^{j}\partial_{\mu}L_{ij}-\frac{1}{2}\gamma_{\mu}\slashed{\partial}L_{ij}\phi^{j} \;,
\nonumber \\
\lambda_{i}&\equiv\gamma^{\mu}J_{\mu i}=\slashed{H}\phi^{j}\varepsilon_{ij}-\slashed{\partial}L_{ij}\phi^{j}\;.
\end{align}
The other components of the current multiplet are obtained by supersymmetry and read
\begin{align}
\label{components_current_1}
t_{\mu}{}^{i}{}_{j}&=\frac{1}{2}H_{\mu}L^{ik}\varepsilon_{kj}+\frac{1}{4}\left(\bar{\phi}^{i}\gamma_{\mu}\phi_{j}-\frac{1}{2}\delta^{i}_{j}\bar{\phi}^{k}\gamma_{\mu}\phi_{k}\right)\;,
\nonumber\\
e^{ij}&=\bar{\phi}^{i}\phi^{j}\;, \quad \tilde{a}_{\mu}=\varepsilon_{ij}\bar{\phi}^{i}\partial_{\mu}\phi^{j}\;, \quad d=\frac{1}{2}L^{ij}L_{ij}\;,
 \nonumber \\
{b}_{\mu\nu}^{-}&=\frac{1}{2}\bar{\phi}^{i}\gamma_{\mu\nu}\phi^{j}\varepsilon_{ij}\;, \quad  c^{ijkl}=L^{(ij}L^{kl)} \;,
\nonumber \\
\Sigma^{ijk}&=L^{(ij}\phi^{k)}\;, \qquad\qquad \xi_{j}=L_{jk}\phi^{k}\;.
\end{align}
Together, eqs. (\ref{components_current}) and (\ref{components_current_1}) form the 48+48 component current multiplet of the rigid N=2 tensor multiplet in four spacetime dimensions. They satisfy the following conservation, reality and trace relations.
\begin{align}\label{relations}
& \partial^{\mu}\theta_{\mu\nu}=0\;, \qquad \sigma=\theta^{\mu}_{\mu}\;, \quad \partial^{\mu}J_{\mu}{}_{i}=0\;, \quad \lambda_{i}=\gamma^{\mu}J_{\mu i}\;, \nonumber \\
& \left(v_{\mu}{}^{i}{}_{j}\right)^{*}=-v_{\mu}{}^{j}{}_{i}\;, \quad v_{\mu}{}^{i}{}_{i}=0\;, \quad \partial_{\mu}v^{\mu}{}^{i}{}_{j}=0\;, \nonumber \\
&\left(t_{\mu}{}^{i}{}_{j}\right)^{*}=-t_{\mu}{}^{j}{}_{i} \;, \quad\;\; t_{\mu}{}^{i}{}_{i}=0\;, \nonumber \\
& \left(c_{ijkl}\right)^{*}\equiv c^{ijkl}=\varepsilon^{im}\varepsilon^{jn}\varepsilon^{kp}\varepsilon^{lq}c_{mnpq}\;,\nonumber \\
& \partial_{\mu}a^{\mu}=0\;, \qquad\quad\;\; \left(a_{\mu}\right)^{*}=a_{\mu}\;,\quad \partial_{\mu}\tilde{a}^{\mu}=0\;, \nonumber \\
& \partial^{\mu}{b}_{\mu\nu}^{-}=\tilde{a}_{\nu}\;, \qquad\qquad\qquad\; \frac{1}{2}\varepsilon_{\mu\nu\rho\sigma}b^{-}{}^{\rho\sigma}=-b_{\mu\nu}^{-}\;,
\end{align}
and transform under supersymmetry as :
\begin{align}\label{current_susy}
\delta\theta_{\mu\nu} &=\bar{\epsilon}^{i}\gamma_{\lambda (\mu}\partial^{\lambda}J_{\nu)i}+\text{h.c.}\;, \nonumber \\
\delta \sigma &= \bar{\epsilon}^i\slashed{\partial}\lambda_i+\text{h.c.}\;, 
 \nonumber \\
 \delta t_{\mu}{}^{i}{}_{j}&=\frac{1}{2}\bar{\epsilon}^{i}J_{\mu}{}_{j}-\frac{1}{2}\bar{\epsilon}^{i}\gamma_{\mu}\lambda_{j}+\frac{1}{4}\varepsilon_{lm}\varepsilon_{jk}\bar{\epsilon}^{l}\slashed{\partial}\left(\gamma_{\mu}\Sigma^{imk}\right)\nonumber 
 \\
 &\quad -\frac{1}{3}\bar{\epsilon}^{i}\gamma_{\mu}\slashed{\partial}\xi_{j}-\frac{1}{3}\bar{\epsilon}^{i}\partial_{\mu}\xi_{j}-\text{(h.c.;traceless)}\;, \nonumber \\
 \delta v_{\mu}{}^{i}{}_{j}&= \frac{1}{2}\bar{\epsilon}^iJ_{\mu j}-\frac{1}{4}\varepsilon_{kj}\varepsilon_{lm}\bar{\epsilon}^l\gamma_\mu\slashed{\partial}\Sigma^{ikm}-\frac{1}{3}\bar{\epsilon}^i\partial_\mu\xi_j \nonumber \\
 &\quad +\frac{1}{4}\varepsilon_{kj}\varepsilon_{lm}\bar{\epsilon}^l\partial_\mu\Sigma^{ikm}+\frac{1}{3}\bar{\epsilon}^i\gamma_\mu\slashed{\partial}\xi_j \nonumber \\
 &\quad -\text{(h.c.;traceless)}\;,  \nonumber \\
\delta a_\mu &= \frac{i}{4}\bar{\epsilon}_k J_\mu^k+\text{h.c.}\;, 
\nonumber \\
\delta b^{-}_{\mu\nu}&=-\varepsilon^{ik}\bar{\epsilon}_i\gamma^\alpha\gamma_{\mu\nu}J_{\alpha k}\;, 
 \nonumber \\
\delta \tilde{a}_\mu &= 2\varepsilon^{ik}\bar{\epsilon}_i\slashed{\partial}J_{\mu k}+\varepsilon^{ik}\bar{\epsilon}_i\gamma_\mu\slashed{\partial}\lambda_k-\varepsilon^{ik}\bar{\epsilon}_{i}\partial_{\mu}\lambda_{k}\;, 
 \nonumber \\
\delta e^{ij}&=-2\bar{\epsilon}_k\slashed{\partial}\Sigma^{ijk}-2\varepsilon^{k(i}\varepsilon^{j)l}\bar{\epsilon}_k\lambda_l-\frac{8}{3}\varepsilon^{k(i}\varepsilon^{j)l}\bar{\epsilon}_k\slashed{\partial}\xi_l \;, 
\nonumber \\
\delta d &= 2 \bar{\epsilon}_i\xi^i+\text{h.c.}\;, 
\nonumber \\
\delta c^{ijkl}&=4\bar{\epsilon}^{(i}\Sigma^{jkl)}+4\varepsilon^{im}\varepsilon^{jn}\varepsilon^{kp}\varepsilon^{lq}\bar{\epsilon}_{(m}\Sigma_{npq)}\;, 
\nonumber \\
 \delta J_{\mu i} &=\theta_{\mu\nu}\gamma^{\nu}\epsilon_{i}+\left(\gamma_{\rho}\gamma_{\mu\sigma}+\gamma_{\mu\sigma}\gamma_{\rho}\right)\partial^{\sigma}v^{\rho}{}^{k}{}_{i}\epsilon_{k}\nonumber \\
 &\quad +\left(\gamma_{\rho}\gamma_{\mu\sigma}+\gamma_{\mu\sigma}\gamma_{\rho}\right)\partial^{\sigma}t^{\rho}{}^{k}{}_{i}\epsilon_{k} -\frac{1}{4}\partial_{\mu}{b}_{\nu\rho}^{-}\gamma^{\nu\rho}\epsilon^{j}\epsilon_{ij}\nonumber \\
 &\quad+\gamma_{\mu\rho}\tilde{a}^{\rho}\epsilon^{j}\varepsilon_{ij}+i\left(\gamma_{\rho}\gamma_{\mu\sigma}-3\gamma_{\mu\sigma}\gamma_{\rho}\right)\partial^{\sigma}a^{\rho}\epsilon_{i}\;, 
\nonumber \\
\delta \lambda_i &= \sigma \epsilon_i+4\slashed{\partial}\slashed{v}^k_i\epsilon_k+4\slashed{\partial}\slashed{t}^k_i\epsilon_k-4\partial_\mu t^{\mu k}_i\epsilon_k-8i\slashed{\partial}\slashed{a}\epsilon_i \nonumber \\
&\quad +3\slashed{\tilde{a}}\epsilon^j\varepsilon_{ij}-\frac{1}{4}\slashed{\partial}(\gamma\cdot b^-)\epsilon^j\varepsilon_{ij}\;, 
 \nonumber \\
\delta \xi_{j} &=-2\slashed{t}^{l}{}_{j}\epsilon_{l}-4\slashed{v}^{l}{}_{j}\epsilon_{l}+6i\slashed{a}\epsilon_{j}-\frac{3}{8}b^{-}.\gamma\epsilon^{m}\varepsilon_{jm}\nonumber \\
&\quad -\frac{1}{2}\varepsilon_{jl}\varepsilon_{km}e^{lk}\epsilon^{m}+\frac{1}{2}\slashed{\partial}d\epsilon_{j}\;, 
\nonumber \\
\delta \Sigma^{ijk} &=-2\varepsilon^{l(i}\slashed{t}^{jk)}\epsilon_{l}+2\varepsilon^{l(i}\slashed{v}^{jk)}\epsilon_{l}+\frac{1}{2}\slashed{\partial}c^{ijkl}\epsilon_{l}-e^{(ij}\epsilon^{k)}\;.
\end{align}
\section{Linearized supergravity multiplet}\label{lin_sugra}
The components of the current multiplet will couple to the linearized components of a supergravity multiplet as shown below in (\ref{coupling_cur_field}). This will fix the basis and normalization of the fields belonging to our linearized gravity multiplet\footnote{If we choose different numerical factors in the coupling or take different linear combination of fields coupling to the currents, we will have a different choice of basis and normalization for the fields which will be related to our choice via field redefinition.}. The invariant action reads:
\begin{widetext}
\begin{align}\label{coupling_cur_field}
S&=\int d^4x \left[\frac{1}{2}\theta^{\mu\nu}h_{\mu\nu}+\sigma\varphi + d\mathcal{D}+\frac{1}{24}c^{ijkl}C_{ijkl}+\frac{1}{4}b^{-}_{\mu\nu}T^{-\mu\nu}-2v_{\mu}{}^i{}_jV^{\mu}{}^j{}_i+e^{ij}E_{ij} +4\,a_\mu A^\mu+\left(t_\mu{}^i{}_j-v_\mu{}^i{}_j\right)S^{\mu j}{}_{i}\right.\nonumber \\
&\quad \quad \quad \quad \quad \left.+2\bar{J}_{\mu i}\psi^{\mu i}+\bar{\lambda}_i\Lambda^i+\bar{\xi}_i\zeta^i+\frac{1}{3}\bar{\Sigma}_{ijk} \Xi^{ijk}+\tilde{a}_{\mu}\tilde{A}^{\mu}+\text{h.c.}\right]\;.
\end{align}
\end{widetext}
Supersymmetry invariance of the above action (\ref{coupling_cur_field}) gives us the linearized transformation of the fields. Due to the constraints (\ref{relations}) satisfied by the currents, the fields coupling to them will have gauge symmetries. For instance, the field $V_{\mu}{}^{i}{}_{j}$ will have an SU(2) gauge symmetry $\delta V_{\mu}{}^{i}{}_{j}\sim\partial_{\mu}\Lambda^{i}{}_{j}$ because it couples to a conserved current (second line of \ref{relations}). As a consequence, we can neglect some pure gauge terms $\sim \bar{\epsilon}^{i}\partial_{\mu}\Lambda_{j}$ that will appear in the supersymmetry variation of $V_{\mu}{}^{i}{}_{j}$. Alternatively one can add terms proportional to $\partial^{\mu}v_{\mu}{}^{j}{}_{i}\epsilon_{j}$ (which is zero due to the relations (\ref{relations})) in the supersymmetry variation of $\lambda_{i}$ in (\ref{current_susy}), which cancels against any pure gauge-like terms of the form shown above in the supersymmetry variation of $V_{\mu}{}^{i}{}_{j}$.

Analogously, due to the last line of (\ref{relations}), the fields $T^{-}_{\mu\nu}$ and $\tilde{A}_{\mu}$ that couple to the current $b^{-}_{\mu\nu}$ and $\tilde{a}_{\mu}$ have a complex vector gauge transformation as shown below.
\begin{align}\label{vec_gauge}
\delta T_{\mu\nu}^{-}&=\partial_{[\mu}\Lambda_{\nu]}-\frac{1}{2}\varepsilon_{\mu\nu\rho\sigma}\partial^{\rho}\Lambda^{\sigma}\;,
\nonumber \\
\delta \tilde{A}_{\mu}&=\Lambda_{\mu}\;.
\end{align}
From the above equation, it is obvious that the field $\tilde{A}_{\mu}$ is a pure gauge under this complex vector gauge transformation and can be gauged away by  using the gauge fixing condition $\tilde{A}_{\mu}=0$. However, we need to compensate the supersymmetry transformation of $\tilde{A}_{\mu}$ by a field-dependent complex vector gauge transformation. This will only affect the supersymmetry transformation of $T^{-}_{\mu\nu}$. After doing the above set of exercises, we obtain the linearized transformation of the fields belonging to the linearized multiplet:
\begin{align}\label{fields_susy_gauge_fixed_redef}
\delta C^{ijkl} &=-2\bar{\epsilon}^{(l}\slashed{\partial}\Xi^{ijk)}-2\varepsilon^{im}\varepsilon^{jn}\varepsilon^{kp}\varepsilon^{lq}\bar{\epsilon}_{(m}\slashed{\partial}\Xi_{npq)}\;, 
 \nonumber \\
\delta \mathcal{D} &= -\frac{1}{4}\bar{\epsilon}_i\slashed{\partial}\chi^i+\text{h.c.}\;, 
 \nonumber \\
\delta h^{\mu\nu}&=2\bar{\epsilon}_i\gamma^{(\mu}\psi^{\nu )i}+\text{h.c.}\;, 
 \nonumber \\
 \delta \varphi &= -\frac{1}{2}\bar{\epsilon}_i\Lambda^i+\text{h.c.} \;, 
 \nonumber \\
 \delta E_{ij}&=\frac{1}{3}\bar{\epsilon}^k\Xi_{ijk}-\frac{1}{2}\bar{\epsilon}^{(l}\zeta^{k)}\varepsilon_{ik}\varepsilon_{jl}\;, 
 \nonumber \\
 \delta V_\mu{}^j{}_i &=-\bar{\epsilon}_i(\gamma_{\rho\sigma}\gamma_\mu+\gamma_\mu\gamma_{\rho\sigma})\partial^\sigma\psi^{\rho j}-2\bar{\epsilon}_i\gamma_\mu\slashed{\partial}\Lambda^j\nonumber \\
 &\quad +\frac{3}{2}\bar{\epsilon}_i\gamma_\mu\zeta^j -\text{(h.c.;traceless)}\;, 
 \nonumber \\
 \delta S_\mu{}^j{}_i &=\bar{\epsilon}_i(\gamma_{\rho\sigma}\gamma_\mu+\gamma_\mu\gamma_{\rho\sigma})\partial^\sigma\psi^\rho{}^j+2\bar{\epsilon}_i\gamma_\mu\slashed{\partial}\Lambda^j-2\bar{\epsilon}_i\partial_\mu\Lambda^j \nonumber \\
 &\quad -\bar{\epsilon}_i\gamma_\mu\zeta^j -\frac{1}{3}\varepsilon^{lj}\varepsilon^{nk}\bar{\epsilon}_n\gamma^\mu\Xi_{ilk}-\text{(h.c.;traceless)}\;, 
 \nonumber \\
 \delta T^{-}{}_{\mu\nu}&=\frac{3}{2}\bar{\epsilon}^i\gamma_{\mu\nu}\zeta^j\epsilon_{ij}-2\bar{\epsilon}^{i}\gamma_{\mu\nu}\slashed{\partial}\Lambda^{j}\epsilon_{ij}\nonumber \\
 &\quad +8\varepsilon_{ij}\bar{\epsilon}^{i}\left(\partial_{[\mu}\psi_{\nu]}^{j}\right)_{|A.S.D}+2\bar{\epsilon}^{i}\gamma_{\mu\nu}\gamma_{\rho\sigma}\partial^{\rho}\psi^{\sigma j}\varepsilon_{ij}\;, 
 \nonumber\\
 \delta A_\mu &=+\frac{i}{4}\bar{\epsilon}_i(\gamma_{\rho\sigma}\gamma_\mu+3\gamma_\mu\gamma_{\rho\sigma})\partial^{\sigma}\psi^\rho{}^i+\frac{3i}{4}\bar{\epsilon}_j\gamma_\mu\zeta^j \nonumber \\
 &\quad -i\bar{\epsilon}_j\gamma_\mu\slashed{\partial}\Lambda^j+\text{h.c.}\;,  \nonumber \\
 \delta \psi^\mu{}^i &=-\frac{1}{2}\gamma_{\lambda\nu}\epsilon^i\partial^\lambda h^{\mu\nu}+V^\mu{}^i{}_j\epsilon^j+{i}A^\mu\epsilon^i \nonumber \\
 &\quad -\frac{1}{8}\varepsilon^{ij}(\gamma\cdot T^{-})\gamma^\mu\epsilon_j\;, 
  \nonumber\\
\delta \Lambda^i&=-2\slashed{\partial}{\varphi}\epsilon^i-\slashed{S}{}^i{}_j\epsilon^j-2\varepsilon^{ik}\varepsilon^{jl}\epsilon_j E_{lk}\;, 
\nonumber \\
\delta \zeta^i &=-\frac{4}{3}\slashed{\partial}\slashed{S}^i{}_j\epsilon^j-\frac{4}{3}\gamma^{\mu\nu}\partial_{\mu}{V}_{\nu}{}^i{}_j\epsilon^j +\frac{8}{3}\varepsilon^{ik}\varepsilon^{lj}\slashed{\partial}{E}_{lk}\epsilon_j \nonumber \\
&\quad -4\mathcal{D}\epsilon^i\;,  \nonumber\\
\delta \Xi_{ijk}&=\frac{3}{2}\varepsilon_{mn}\varepsilon_{lp}\left[\partial_{\mu}S^\mu{}^l{}_{(i}\delta^n_j\delta^p_{k)}-2\gamma^{\mu\nu}\partial_{\mu}S_{\nu}{}^l{}_{(i}\delta^n_j\delta^p_{k)}\right. \nonumber \\
&\quad \quad \quad \quad \quad \quad \left.-2\gamma^{\mu\nu}\partial_{\mu}V_{\nu}{}^l{}_{(i}\delta^n_j\delta^p_{k)} \right]\epsilon^m
\nonumber \\
& \quad +6\slashed{\partial}E_{(ij}\epsilon_{k)}-C_{ijkl}\epsilon^l\;.
\end{align}
In the above variations, we observe that the derivative of the gravitino $\psi_{\mu}^{i}$ appears in antisymmetric form $\sim\partial_{[\mu}\psi_{\nu]}^{i}$, in the supersymmetry variation of some of the fields like $V_{\mu}{}^{i}{}_{j}$, $S_{\mu}{}^{i}{}_{j}$, $T^{-}_{\mu\nu}$ and $A_{\mu}$. This is important because derivatives of gauge fields should be traded off for the corresponding curvature (R(Q) in this case) and it can only happen if the derivatives appear in an anti-symmetric form. But before trading off $\partial \psi$ in terms of R(Q) in the above variations, we recall that the curvature R(Q) will contain a term proportional to the S-gauge field $\phi_{\mu}^{i}$ at the linearized level. This happens because the gravitino $\psi_{\mu}^{i}$ transforms under S-supersymmetry which is known to be $\delta_S \psi_{\mu}{}^{i}=-\gamma_{\mu}\eta^{i}$ from the superconformal algebra. Hence, the linearized form of the curvature $R(Q)_{\mu\nu}^{i}$, is given as 
\begin{align}\label{RQ}
R(Q)_{\mu\nu}^{i}=2\partial_{[\mu}\psi_{\nu]}^{i}-\gamma_{[\mu}\phi_{\nu]}^{i}\;, 
\end{align}
where $\phi_{\mu}^{i}$ is the gauge field corresponding to S-supersymmetry. Using (\ref{RQ}) one can trade off the antisymmetric derivative on the gravitino for the curvature R(Q) in the supersymmetry variation of the above-mentioned fields. However, we also need to impose the curvature constraint on R(Q).  The conventional constraint for R(Q)  (\ref{constraints}) involves the standard Weyl multiplet field $\chi^{i}$. However, at this point, since we have not yet identified the field $\chi^{i}$, which would be some linear combination of the fields appearing in (\ref{fields_susy_gauge_fixed_redef}), it is difficult to implement the conventional constraint for the curvature R(Q) from (\ref{constraints}). We will instead impose the following unconventional constraint on R(Q)\footnote{The conventional and the unconventional constraints are related by a shift in the S-gauge field by a term proportional to the auxiliary fermion $\chi^{i}$. Once we correctly identify the $\chi^{i}$, we can go back to the conventional constraint by making an appropriate shift in the S-gauge field.}
\begin{align}\label{unconRQ}
\gamma^{\mu}R(Q)_{\mu\nu}^{i}=0\;, 
\end{align}
which, at the linearized level, gives:
\begin{align}\label{linRQ}
\gamma^{\mu\nu}\partial_{\mu}\psi_{\nu}^{i}=-\frac{3}{2}\gamma^{\nu}\phi_{\nu}^{i}\;.
\end{align}
We can see that, when we try to re-write $\partial_{[\mu}\psi_{\nu]}^{i}$ in terms of R(Q) and use the above constraint in the linearized transformation (\ref{fields_susy_gauge_fixed_redef}), bare S-gauge field $\phi_{\mu}^{i}$ will pop out as shown below. 

\begin{align}\label{S-susyfixing}
\delta V_{\mu}{}^{i}{}_{j}&= 2\bar{\epsilon}_{j}\phi_{\mu}^{i}-2\bar{\epsilon}_{j}\gamma_{\mu}\gamma\cdot\phi^{i}-2\bar{\epsilon}_{j}\slashed{\partial}\Lambda^{i}+\frac{3}{2}\bar{\epsilon}_{j}\gamma_{\mu}\zeta^{i}\nonumber \\
& \quad-(\text{h.c.;traceless})\;, 
\nonumber \\
\delta A_{\mu}&=\frac{i}{2}\bar{\epsilon}_{i}\phi_{\mu}^{i}-i\bar{\epsilon}_{j}\gamma_{\mu}\gamma\cdot\phi^{j}+\frac{3i}{4}\bar{\epsilon}_j\gamma_\mu\zeta^j -i\bar{\epsilon}_j\gamma_\mu\slashed{\partial}\Lambda^j\nonumber \\
&\quad +\text{h.c.}\;, 
 \nonumber \\
\delta T_{\mu\nu}^{-} &= 4\varepsilon_{ij}\bar{\epsilon}^{i}R(Q)_{\mu\nu}^{j}-2\bar{\epsilon}^{i}\gamma_{\mu\nu}\slashed{\partial}\Lambda^{j}\varepsilon_{ij}-2 \bar{\epsilon}^{i}\gamma_{\mu\nu}\gamma\cdot\phi^{j}\varepsilon_{ij}\nonumber \\
&\quad +\frac{3}{2}\bar{\epsilon}^i\gamma_{\mu\nu}\zeta^j\epsilon_{ij}\;, 
\nonumber \\
\delta S_{\mu}{}^{i}{}_{j} &=2\bar{\epsilon}_i\gamma_\mu\slashed{\partial}\Lambda^j-2\bar{\epsilon}_i\partial_\mu\Lambda^j +2\bar{\epsilon}_i\gamma_\mu\gamma\cdot\phi^{j}-2\bar{\epsilon}_i\gamma_\mu\phi^j \nonumber \\
&\quad -\bar{\epsilon}_i\gamma_\mu\zeta^j -\frac{1}{3}\varepsilon^{lj}\varepsilon^{nk}\bar{\epsilon}_n\gamma^\mu\Xi_{ilk}-\text{(h.c.;traceless)}\;.
\end{align}
Among the fields that appear above in (\ref{S-susyfixing}), the fields like $S_{\mu}{}^{i}{}_{j}$ and $T^{-}_{\mu\nu}$ do not couple to any conserved current, and as a result they will not correspond to any gauge fields of the multiplet. Rather they will be the auxiliary fields. Such fields should not have any bare S-gauge field in their transformation rules, which should be either be inside a covariant derivative or a curvature. Hence, terms involving bare S-gauge field that appears in (\ref{S-susyfixing}) for $T_{\mu\nu}^{-}$ and $S_{\mu}{}^{i}{}_{j}$, should be absorbed into the covariant derivative of $\Lambda^{i}$. This can be done, if we let $\Lambda^{i}$ to transform under S-supersymmetry as 
\begin{align}\label{S-susy-lambda}
\delta_{S}\Lambda^{i}=-2\eta^{i}\;.
\end{align}
As a consequence, the covariant derivative of $\Lambda^{i}$ at the linearized level is as shown below:
\begin{align}\label{covderlambda}
D_{\mu}\Lambda^{i}=\partial_{\mu}\Lambda^{i}+\phi_{\mu}^{i}\;.
\end{align}
In terms of the covariant derivative on $\Lambda^{i}$, the supersymmetry variation (\ref{S-susyfixing}) takes the following form
\begin{align}\label{S-susyfixing1}
\delta V_{\mu}{}^{i}{}_{j}&= 2\bar{\epsilon}_{j}\phi_{\mu}^{i}-2\bar{\epsilon}_{j}\slashed{D}\Lambda^{i}+\frac{3}{2}\bar{\epsilon}_{j}\gamma_{\mu}\zeta^{i}-(\text{h.c.;traceless})\;, 
\nonumber \\
\delta A_{\mu}&=\frac{i}{2}\bar{\epsilon}_{i}\phi_{\mu}^{i}+\frac{3i}{4}\bar{\epsilon}_j\gamma_\mu\zeta^j -i\bar{\epsilon}_j\gamma_\mu\slashed{D}\Lambda^j+\text{h.c.}\;,  \nonumber \\
\delta T_{\mu\nu}^{-} &= 4\varepsilon_{ij}\bar{\epsilon}^{i}R(Q)_{\mu\nu}^{j}-2\bar{\epsilon}^{i}\gamma_{\mu\nu}\slashed{D}\Lambda^{j}\varepsilon_{ij}+\frac{3}{2}\bar{\epsilon}^i\gamma_{\mu\nu}\zeta^j\varepsilon_{ij}\;, 
\nonumber \\
\delta S_{\mu}{}^{i}{}_{j} &=2\bar{\epsilon}_i\gamma_\mu\slashed{D}\Lambda^j-2\bar{\epsilon}_i D_\mu\Lambda^j -\bar{\epsilon}_i\gamma_\mu\zeta^j \nonumber \\
&\quad -\frac{1}{3}\varepsilon^{lj}\varepsilon^{nk}\bar{\epsilon}_n\gamma^\mu\Xi_{ilk}-\text{(h.c.;traceless)}\;.
\end{align}
By looking at the transformation of the R-symmetry gauge fields $V_{\mu}{}^{i}{}_{j}$ and $A_{\mu}$, we find that the standard Weyl multiplet $\chi^{i}$ can be identified with the following linear combination at the linearized level\footnote{Here, we need to keep in mind that we are working with the unconventional constraint and hence the variation of $V_{\mu}{}^{i}{}_{j}$ and $A_{\mu}$ should be compared with the results known in the literature after shifting the S-gauge field so that we can correctly identify the $\chi$ field (see for instance \cite{Butter:2017pbp}).} .
\begin{equation}\label{fieldredef}
{\chi}^{i}=-\frac{3}{4}\zeta^{i}+\slashed{D}\Lambda^{j}\;.
\end{equation}
By making a supersymmetry transformation on the above, we find that the following linear combination is identified with the standard-Weyl multiplet field ``$D$'' at the linearized level.
\begin{align}\label{fieldredef1}
{D}&= 3\mathcal{D} -2\partial^{2}\varphi\;.
\end{align}
Finally, we also redefine the scalar field $\varphi$ as \footnote{The redefinitions are redundant at the linearized level, but important at the non-linear level.}
\begin{align}\label{scalarredef}
\phi&=e^{\varphi}\;.
\end{align} 
Having identified the standard-Weyl multiplet field $\chi^{i}$, we would prefer going back to the conventional constraint by making a field redefinition on the S-gauge field: $\phi_{\mu}^{i}\to \phi_{\mu}^{i}-\frac{1}{2}\gamma_{\mu}\chi^{i}$. In terms of the above redefined fields ((\ref{fieldredef})-(\ref{scalarredef}) as well as the redefined S-gauge field), we see that the linearized components $h_{\mu\nu}$, $V_{\mu}{}^{i}{}_{j}$, $A_{\mu}$, $T^{-}_{\mu\nu}$, ${D}$, $\psi_{\mu}{}^{i}$, ${\chi}^{i}$ decouples from the rest and they transform, under Q-supersymmetry, exactly like the linearization of the standard Weyl multiplet known in the literature in the conventional constraints. The full supersymmetry transformation of the above fields including the S-supersymmetry transformation follows from the superconformal algebra (\ref{algebra}) and is already known. The remaining 24+24 components $\phi$, $E_{ij}$, $S_{\mu}{}^{i}{}_{j}$, $C_{ijkl}$, $\Lambda^{i}$, $\Xi_{ijk}$ transform as given below.
\begin{align}\label{Lin_mult}
\delta \phi &= -\frac{\phi}{2}\bar{\epsilon}^{i}\Lambda_{i}+\text{h.c.}\;,
 \nonumber \\
\delta \Lambda^{i}&=-2\slashed{P}\epsilon^{i}-\left(\slashed{S}^{i}{}_{j}\epsilon^{j}+2\varepsilon^{ik}\varepsilon^{jl}\epsilon_{j}E_{lk}\right) -2\eta^{i}\;,
\nonumber \\
\delta S_{a}{}^{i}{}_{j}&=\bar{\epsilon}_{j}\gamma_{a}\chi^{i}+\frac{2}{3}\bar{\epsilon}_{j}\gamma_{a}\slashed{D}\Lambda^{i}-2\bar{\epsilon}_{j}D_{a}\Lambda^{i}\nonumber \\
&\quad -\frac{1}{3}\varepsilon^{li}\varepsilon^{nk}\bar{\epsilon}_{n}\gamma_{a}\Xi_{ljk}-\text{(h.c.;traceless)}\;,\nonumber \\
\delta E_{ij}&=2\bar{\epsilon}^{(l}\chi^{k)}\varepsilon_{ik}\varepsilon_{jl}-\frac{2}{3}\bar{\epsilon}^{(l}\slashed{D}\Lambda^{k)}\varepsilon_{ik}\varepsilon_{jl}+\frac{1}{3}\bar{\epsilon}^{k}\Xi_{ijk}\;, \nonumber \\
\delta \Xi_{ijk}&=\frac{3}{2}\varepsilon_{mn}\varepsilon_{lp}\left[\partial_{\mu}S^\mu{}^l{}_{(i}\delta^n_j\delta^p_{k)}-2\gamma^{\mu\nu}\partial_{\mu}S_{\nu}{}^l{}_{(i}\delta^n_j\delta^p_{k)}\right. \nonumber \\
&\quad \quad \quad \quad \quad \quad \left. -\gamma\cdot R(V)^l{}_{(i}\delta^n_j\delta^p_{k)} \right]\epsilon^m\nonumber \\
&\quad +6\slashed{\partial}E_{(ij}\epsilon_{k)}-C_{ijkl}\epsilon^l\;,
\nonumber \\
\delta C^{ijkl}&=-2\bar{\epsilon}^{(l}\slashed{\partial}\Xi^{ijk)}-2\varepsilon^{im}\varepsilon^{jn}\varepsilon^{kp}\varepsilon^{lq}\bar{\epsilon}_{(m}\slashed{\partial}\Xi_{npq)}\;.
\end{align}
Here, we have defined $P_{a}=\phi^{-1}\partial_{a}\phi$ and $R(V)_{\mu\nu}{}^{i}{}_{j}=2\partial_{[\mu}V_{\nu]}{}^{i}{}_{j}$. The above result (\ref{Lin_mult}) is a linearization of our final result (\ref{Susy-transf})\footnote{We have partial non-linear results (see $\delta\phi$ and first term of $\delta\Lambda^{i}$) because we worked with the field $\phi$ instead of $\varphi$ which are related by an exponentiation as given in (\ref{scalarredef}).}. The above linearized transformation gives us the Weyl and chiral weights of the field (See Table-\ref{Table-New}). Because of its S-susy transformation, $\Lambda^{i}$ will have the same chiral and Weyl weight as $\eta^{i}$. Further,  we obtain the weights of $E_{ij}$, $S_{a}{}^{i}{}_{j}$, $C_{ijkl}$ and $\Xi_{ijk}$ by looking at the above susy transformation (\ref{Lin_mult}) and the knowledge of the weights of the Q-susy parameter $\epsilon^{i}$. The chiral weight of $\phi$ has to be zero because it is a real scalar field while its Weyl weight is obtained by acting the commutator $[\delta_S,\delta_Q]$ on it. We present a schematic Table-\ref{Table-New} with the full field content of the multiplet, along with their SU(2) irreps, Weyl and chiral weights as well as chirality for fermions.
\begin{table}\caption{}\label{Table-New}
\begin{center}
  \begin{tabular}{ | p{1cm}|p{2cm}|p{1cm}|p{1cm}|p{2cm}| }
    \hline
    Field & SU(2) Irreps & Weyl weight (w) & Chiral weight (c) & Chirality (Fermions) \\ \hline
   $\phi$ & $\bf{1}$ & 1 & 0 & -- \\ \hline
   $S_{a}{}^{i}{}_{j}$ & $\bf{3}$ & 1 & 0 & -- \\ \hline
   $E_{ij}$ & $\bf{3}$ & 1 & -1 & -- \\ \hline
   $C_{ijkl}$ & $\bf{5}$ & 2 & 0 & -- \\ \hline
   $\Lambda_{i}$ & $\bf{2}$ & 1/2 & 1/2 & +1 \\ \hline
   $\Xi_{ijk}$ & $\bf{4}$ & 3/2 & -1/2 & +1 \\ \hline
  \end{tabular}
\end{center}
\end{table}
The chirality and the SU(2) irreps of the fields are obvious from the currents they couple to. In addition the field $S_{a}{}^{i}{}_{j}$ and $C_{ijkl}$ satisfy the following reality conditions which again follow from the reality conditions satisfied by the corresponding currents.
\begin{align}\label{Sreality}
(S_{a}{}^{i}{}_{j})^{*}&\equiv S_{a}{}_{i}{}^{j}=-S_{a}{}^{j}{}_{i}\;,  \nonumber \\
(C_{ijkl})^{*}&\equiv C^{ijkl}=\varepsilon^{im}\varepsilon^{jn}\varepsilon^{kp}\varepsilon^{lq}C_{mnpq}\;.
\end{align}

In the next section, we will show the fully non-linear result (\ref{Susy-transf}), obtained by applying the supersymmetry algebra (\ref{algebra}) on the linearized result (\ref{Lin_mult}). We will also assume, without loss of generality, that the S-transformation of $S_{a}{}^{i}{}_{j}$, $E_{ij}$, $C_{ijkl}$ and $\Xi_{ijk}$ is zero. This is because, given its S-susy transformation, $\Lambda^i$ can be straightforwardly used as a compensator to give S-susy invariant fields.

\section{The real scalar multiplet}\label{result}
Completing the linearized analysis of the previous section gives us the full supersymmetry transformation of the real scalar multiplet. All the field components are invariant under special conformal transformation\footnote{ This follows in a straightforward way from the $[\delta_S,\delta_S]$ commutators (\ref{algebra}).}, and they transform under Q- and S-supersymmetry as shown below\footnote{ We use the usual conventions for symmetrized non-consecutive indices, i.e. the indices that are not symmetrized are surrounded by vertical lines. For instance, we will write a multi-indices tensor product $A_{i}B_{jkl}$, symmetrized in $(ijl)$ as $A_{(i}B_{j|k|l)}$. }.
\begin{widetext}
\begin{align}\label{Susy-transf}
 \delta \phi &= -\frac{\phi}{2}\bar{\epsilon}^{i}\Lambda_{i}+\text{h.c.}\;, \nonumber \\
 \delta \Lambda^{i}&=-2\slashed{P}\epsilon^{i}-\left(\slashed{S}^{i}{}_{j}\epsilon^{j}+2\varepsilon^{ik}\varepsilon^{jl}\epsilon_{j}E_{lk}\right)-\frac{1}{2}\bar{\Lambda}^{i}\Lambda^{j}\epsilon_{j}-\frac{1}{4}\bar{\Lambda}^{j}\gamma_{a}\Lambda_{j}\gamma^{a}\epsilon^{i}+\frac{1}{8}\bar{\Lambda}^{i}\gamma_{ab}\Lambda^{j}\gamma^{ab}\epsilon_{j}-2\eta^{i}\;,\nonumber\\
  \delta S_{a}{}^{i}{}_{j}&=\bar{\epsilon}_{j}\gamma_{a}\chi^{i}+\frac{2}{3}\bar{\epsilon}_{j}\gamma_{a}\slashed{D}\Lambda^{i}-2\bar{\epsilon}_{j}D_{a}\Lambda^{i}-\frac{1}{3}\varepsilon^{li}\varepsilon^{nk}\bar{\epsilon}_{n}\gamma_{a}\Xi_{ljk}+\frac{1}{24}\bar{\epsilon}_{j}\gamma_{a}\gamma.T^{-}\Lambda_{k}\varepsilon^{ik}-\frac{1}{3}\bar{\epsilon}_{j}\gamma_{a}\Lambda_{k}E_{lm}\varepsilon^{il}\varepsilon^{km}-\frac{2}{3}\bar{\epsilon}^{i}\gamma_{a}\slashed{S}^{k}{}_{j}\Lambda_{k} \nonumber\\
 &\quad -\frac{1}{2}\bar{\epsilon}^{i}\slashed{S}^{k}{}_{j}\gamma_{a}\Lambda_{k}+\frac{1}{2}\bar{\epsilon}^{k}\gamma_{a}\slashed{S}^{i}{}_{j}\Lambda_{k} -\frac{2}{3}\bar{\epsilon}^{i}\gamma_{a}\slashed{P}\Lambda_{j}-\bar{\epsilon}^{i}\slashed{P}\gamma_{a}\Lambda_{j}-\frac{1}{24}\bar{\Lambda}^{i}\Lambda^{k}\bar{\epsilon}_{j}\gamma_{a}\Lambda_{k}-\frac{1}{32}\bar{\Lambda}^{i}\gamma_{bc}\Lambda^{k}\bar{\epsilon}_{j}\gamma^{bc}\gamma_{a}\Lambda_{k}\nonumber \\
 &\quad -\text{(h.c.;traceless)}\;,\nonumber\\
  \delta E_{ij}&=2\bar{\epsilon}^{(l}\chi^{k)}\varepsilon_{ik}\varepsilon_{jl}-\frac{2}{3}\bar{\epsilon}^{(l}\slashed{D}\Lambda^{k)}\varepsilon_{ik}\varepsilon_{jl}+\frac{1}{3}\bar{\epsilon}^{k}\Xi_{ijk}-\frac{1}{12}\bar{\epsilon}^{k}\gamma.T^{-}\Lambda_{(i}\varepsilon_{j)k}+\frac{2}{3}\bar{\epsilon}^{k}\Lambda_{(i}E_{j)k} -2\bar{\epsilon}_{(i}\Lambda^{k}E_{j)k}-\frac{2}{3}\bar{\epsilon}^{k}\Lambda_{k}E_{ij} \nonumber\\
 & \quad+\bar{\epsilon}_{k}\Lambda^{k}E_{ij}-\frac{1}{3}\bar{\epsilon}^{(l}\slashed{S}^{m)}{}_{k}\Lambda^{k}\varepsilon_{il}\varepsilon_{jm}-\frac{2}{3}\bar{\epsilon}^{(k}\slashed{P}\Lambda^{l)}\varepsilon_{ik}\varepsilon_{jl}-\frac{1}{12}\bar{\epsilon}^{(l}\gamma_{a}\Lambda^{k)}\bar{\Lambda}^{m}\gamma^{a}\Lambda_{m}\varepsilon_{il}\varepsilon_{jk}\;, 
 \nonumber \\
 \delta \Xi_{ijk}&=\frac{3}{2}\epsilon_{mn}\epsilon_{lp}\left[D_{a}S^a{}^l{}_{(i}\delta^n_j\delta^p_{k)}-2\gamma^{ab}D_{a}S_{b}{}^l{}_{(i}\delta^n_j\delta^p_{k)}-\gamma.R(V)^l{}_{(i}\delta^n_j\delta^p_{k)} \right]\epsilon^m+{6}\slashed{D}E_{(ij}\epsilon_{k)} -C_{ijkl}\epsilon^l
 \nonumber \\
  &\quad-6E^{ln}E_{m(i}\varepsilon_{j|l|}\varepsilon_{k)n}\epsilon^{m} -6\slashed{P}E_{(ij}\epsilon_{k)}+3\slashed{S}^{m}{}_{(i}E_{jk)}\epsilon_{m}-6\slashed{S}^{m}{}_{(i}E_{j|m|}\epsilon_{k)}+3S^{a}{}^{m}{}_{(i}S^{b}{}^{n}{}_{j}\varepsilon_{k)m}\varepsilon_{ln}\gamma_{ab}\epsilon^{l}   
   \nonumber\\
 &\quad + 3 P^{a}S_{a}{}^{l}{}_{(i}\varepsilon_{j|l|}\varepsilon_{k)m}\epsilon^{m}-\frac{3}{4}\gamma.T^{-}E^{lm}\epsilon^{n}\varepsilon_{(i|l|}\varepsilon_{j|m|}\varepsilon_{k)n}+\bar{\Lambda}^{l}\gamma_{a}\slashed{D}\Lambda^{m}\gamma^{a}\epsilon_{(i}\varepsilon_{j|l|}\varepsilon_{k)m}+\frac{1}{4}\bar{\Lambda}^{l}\slashed{D}\Lambda_{(i}\epsilon^{m}\varepsilon_{j|l|}\varepsilon_{k)m}
 \nonumber \\
 & \quad-{\frac{3}{8}}\bar{\Lambda}^{l}\slashed{D}\gamma_{ab}\Lambda_{(i}\gamma^{ab}\epsilon^{m}\varepsilon_{j|l|}\varepsilon_{k)m}-{\frac{1}{4}}\bar{\Lambda}_{(i}\slashed{D}\Lambda^{l}\epsilon^{m}\varepsilon_{j|l|}\varepsilon_{k)m} +{\frac{1}{8}}\bar{\Lambda}_{(i}\gamma_{ab}\slashed{D}\Lambda^{l}\gamma^{ab}\epsilon^{m}\varepsilon_{j|l|}\varepsilon_{k)m} 
 \nonumber \\
  &\quad-\frac{3}{2}\bar{\Lambda}_{(i}R(Q)_{ab}{}^{l}\gamma^{ab}\epsilon^{m}\varepsilon_{j|l|}\varepsilon_{k)m}+\frac{1}{2}\bar{\Lambda}^{p}\Xi^{lmn}\epsilon^{q}\varepsilon_{il}\varepsilon_{jm}\varepsilon_{kn}\varepsilon_{pq}-\frac{1}{2}\bar{\Lambda}^{(m}\Xi^{np)l}\epsilon^{q}\varepsilon_{im}\varepsilon_{jn}\varepsilon_{kp}\varepsilon_{lq}-\frac{1}{8}\bar{\Lambda}_{l}\gamma_{ab}\Xi_{ijk}\gamma^{ab}\epsilon^{l}
  \nonumber \\
  &\quad -\bar{\Lambda}_{(i}\Xi_{jk)l}\epsilon^{l}+\frac{1}{8}\bar{\Lambda}_{(i}\gamma_{ab}\Xi_{jk)l}\gamma^{ab}\epsilon^{l} -\frac{1}{2}\bar{\Lambda}^{l}\gamma_{a}\Xi_{ijk}\gamma^{a}\epsilon_{l}+\bar{\Lambda}^{l}\gamma_{a}\Xi_{l(ij}\gamma^{a}\epsilon_{k)}-\frac{3}{2}\bar{\Lambda}_{(i}\chi^{l}\epsilon^{m}\varepsilon_{j|l|}\varepsilon_{k)m}
  \nonumber \\
  & \quad -\frac{3}{2}\bar{\Lambda}_{(i}\gamma_{ab}\chi^{l}\gamma^{ab}\epsilon^{m}\varepsilon_{j|l|}\varepsilon_{k)m}+\frac{3}{2}\bar{\Lambda}^{l}\chi_{(i}\epsilon^{m}\varepsilon_{j|l|}\varepsilon_{k)m}-3\bar{\Lambda}^{l}\gamma_{a}\chi^{m}\gamma^{a}\epsilon_{(i}\varepsilon_{j|l|}\varepsilon_{k)m}+\frac{1}{4}\varepsilon_{l(k}\bar{\Lambda}_{i}\Lambda_{j)}\gamma.T^{-}\epsilon^{l} \nonumber \\
  &\quad -\frac{1}{8}\varepsilon_{l(k}\bar{\Lambda}^{l}\gamma_{a}\Lambda_{i}\gamma.T^{-}\gamma^{a}\epsilon_{j)}+\frac{1}{2}\bar{\Lambda}^{(l}\Lambda^{m}E^{n)p}\epsilon^{q}\varepsilon_{il}\varepsilon_{jm}\varepsilon_{kn}\varepsilon_{pq}-\frac{1}{2}\bar{\Lambda}^{p}\Lambda^{(l}E^{mn)}\epsilon^{q}\varepsilon_{il}\varepsilon_{jm}\varepsilon_{kn}\varepsilon_{pq}+\frac{1}{2}\bar{\Lambda}_{l}\Lambda_{(i}E_{jk)}\epsilon^{l} 
  \nonumber \\
  &\quad -\frac{1}{2}\bar{\Lambda}_{(i}\Lambda_{j}E_{k)l}\epsilon^{l}+\frac{1}{4}\bar{\Lambda}_{l}\gamma_{ab}\Lambda_{(i}E_{jk)}\gamma^{ab}\epsilon^{l} +\frac{1}{4}\bar{\Lambda}^{l}\gamma_{a}\Lambda_{l}\gamma^{a}\epsilon_{(i}E_{jk)}-\bar{\Lambda}^{l}\gamma_{a}\Lambda_{(i}\gamma^{a}\epsilon_{j}E_{k)l}-\frac{1}{2}\bar{\Lambda}^{l}\slashed{S}^{n}{}_{l}\Lambda_{(i}\epsilon^{m}\varepsilon_{j|n|}\varepsilon_{k)m}
  \nonumber \\
  &\quad +\frac{1}{4}\varepsilon_{(i|n|}\varepsilon_{j|m|}\bar{\Lambda}^{l}\slashed{S}^{n}{}_{k)}\Lambda_{l}\epsilon^{m}-\frac{1}{8}\bar{\Lambda}^{l}\slashed{S}^{n}{}_{l}\gamma_{ab}\Lambda_{(i}\gamma^{ab}\epsilon^{m}\varepsilon_{j|n|}\varepsilon_{k)m}+\frac{3}{16}\varepsilon_{(i|n|}\varepsilon_{j|m|}\bar{\Lambda}^{l}\slashed{S}^{n}{}_{k)}\gamma^{ab}\Lambda_{l}\gamma_{ab}\epsilon^{m}
  \nonumber \\
 &\quad  +\frac{1}{2}\bar{\Lambda}^{n}\Lambda^{m}\slashed{S}^{l}{}_{(i}\epsilon_{j}\varepsilon_{k)m}\varepsilon_{nl} +\frac{1}{2}\bar{\Lambda}^{l}\slashed{P}\Lambda_{(i}\epsilon^{m}\varepsilon_{j|l|}\varepsilon_{k)m}+\frac{1}{16}\bar{\Lambda}^{l}\gamma_{ab}\Lambda^{m}\slashed{S}^{n}{}_{(i}\gamma^{ab}\epsilon_{j}\varepsilon_{k)n}\varepsilon_{lm}+\frac{1}{2}\bar{\Lambda}^{l}\slashed{P}\gamma_{ab}\Lambda_{(i}\gamma^{ab}\epsilon^{m}\varepsilon_{j|l|}\varepsilon_{k)m}
 \nonumber \\
  &\quad +\bar{\Lambda}^{l}\Lambda^{m}\slashed{P}\epsilon_{(i}\varepsilon_{j|l|}\varepsilon_{k)m}+\frac{1}{8}\bar{\Lambda}^{m}\Lambda^{n}\bar{\Lambda}^{l}\gamma_{a}\Lambda_{l}\gamma^{a}\epsilon_{(i}\varepsilon_{j|m|}\varepsilon_{k)n} -\frac{1}{16}\bar{\Lambda}^{l}\Lambda^{m}\bar{\Lambda}_{n}\gamma_{ab}\Lambda_{(i}\gamma^{ab}\epsilon^{n}\varepsilon_{j|l|}\varepsilon_{k)m}\;,
  \nonumber \\
 \delta C_{ijkl}&=\bar{\epsilon}_{(i}\Gamma_{jkl)}+\varepsilon_{ip}\varepsilon_{jq}\varepsilon_{kr}\varepsilon_{ls}\bar{\epsilon}^{(p}\Gamma^{qrs)}+\left(\bar{\epsilon}_{m}\Lambda^{m}+\bar{\epsilon}^{m}\Lambda_{m}\right)C_{ijkl}\;.
 \end{align}
In the above, the composite field $P_{a}\equiv \phi^{-1}D_{a}\phi$ and $\Gamma_{ijk}$ is defined as:
 \begin{align}\label{Gammadef}
 \Gamma_{ijk}&=-2\slashed{D}\Xi_{ijk}-3D_{a}S^{a}{}^{n}{}_{(i}\Lambda^{m}\varepsilon_{j|n|}\varepsilon_{k)m}+6\slashed{D}E_{(ij}\Lambda_{k)}-2C_{ijkl}\Lambda^{l}+2\slashed{D}\Lambda_{(i}E_{jk)}+2D_a\Lambda^l S^a{}^n{}_{(i}\varepsilon_{j|n|}\varepsilon_{k)l}
 \nonumber \\
 &\quad -2\gamma^{ab}D_a\Lambda^l S_b{}^n{}_{(i}\varepsilon_{j|n|}\varepsilon_{k)l} -4\Xi^{lmn}E_{l(i}\varepsilon_{j|m|}\varepsilon_{k)n}+2\slashed{S}^{l}{}_{(i}\Xi_{jk)l}+12 \chi_{(i}E_{jk)}-6\slashed{S}^{l}{}_{(i}\chi^{m}\varepsilon_{j|l|}\varepsilon_{k)m} -4\slashed{P}E_{(ij}\Lambda_{k)}
 \nonumber \\
 &\quad  -4P_aS^a{}^n{}_{(i}\varepsilon_{j|n|}\varepsilon_{k)l}\Lambda^l -2P_a\gamma^{ab}S_b{}^n{}_{(i}\varepsilon_{j|n|}\varepsilon_{k)l}\Lambda^l-2\slashed{S}^l{}_{(i}E_{jk)}\Lambda_l-2\slashed{S}^l{}_{(i}E_{j|l|}\Lambda_{k)}-4E_{(ij}\varepsilon_{k)m}\varepsilon_{ln}E^{mn}\Lambda^l 
  \nonumber \\
 &\quad +12\varepsilon_{(i|m|}\varepsilon_{j|n|}E_{k)l}E^{mn}\Lambda^l+S_a{}^m{}_{(i}S^a{}^n{}_j\varepsilon_{k)m}\varepsilon_{ln}\Lambda^l+\gamma_{ab}S^{a m}{}_{(i}S^b{}^n{}_j\varepsilon_{k)m}\varepsilon_{ln}\Lambda^l +\frac{1}{2}\gamma\cdot T^+ E_{(ij}\varepsilon_{k)l}\Lambda^l
 \nonumber \\
 &\quad  +\frac{1}{4}\slashed{S}^l{}_{(i}\gamma\cdot T^-\Lambda_{j}\varepsilon_{k)l}+\frac{1}{2}\bar{\Lambda}^l\Lambda^m\slashed{D}\Lambda_{(i}\varepsilon_{j|l|}\varepsilon_{k)m} +\frac{5}{4}\bar{\Lambda}^l\gamma^a\Lambda_{(i}\varepsilon_{j|l|}\varepsilon_{k)m}D_a\Lambda^m+\frac{5}{16}\bar{\Lambda}^l\gamma^{bc}\gamma^a\Lambda_{(i}\varepsilon_{j|l|}\varepsilon_{k)m}\gamma_{bc}D_a\Lambda^m
 \nonumber \\
 &\quad +\frac{3}{2}\bar{\Lambda}^l\Lambda^m\chi_{(i}\varepsilon_{j|l|}\varepsilon_{k)m}-\frac{3}{2}\bar{\Lambda}^l\gamma^a\Lambda_{(i}\varepsilon_{j|l|}\varepsilon_{k)m}\gamma_a\chi^m +\frac{1}{2}\bar{\Lambda}^{q}\Lambda^{(p}\Xi^{ml)n}\varepsilon_{ip}\varepsilon_{jm}\varepsilon_{kl}\varepsilon_{qn} +\frac{1}{2}\bar{\Lambda}^l\gamma_a\Lambda_{(i}\gamma^a\Xi_{jk)l}\nonumber \\
 &\quad +\frac{3}{2}\bar{\Lambda}^l\Lambda^m\slashed{P}\Lambda_{(i}\varepsilon_{j|l|}\varepsilon_{k)m}-\frac{1}{4}\bar{\Lambda}^l\Lambda^m \slashed{S}^n{}_{(i}\varepsilon_{j|n|}\varepsilon_{k)m}\Lambda_l -\frac{1}{32}\varepsilon_{mn}\bar{\Lambda}^m\gamma^{ab}\Lambda^n\gamma_{ab}\slashed{S}^l{}_{(i}\Lambda_j\varepsilon_{k)l}
 \nonumber \\
 &\quad -\frac{1}{2}\bar{\Lambda}^l\Lambda^{(m}E^{pq)}\Lambda^n\varepsilon_{ip}\varepsilon_{jq}\varepsilon_{km}\varepsilon_{ln}-\bar{\Lambda}_{(i}\Lambda_j E_{k)l}\Lambda^l-\bar{\Lambda}_l\Lambda_{(i}E_{jk)}\Lambda^l+\frac{1}{8}\bar{\Lambda}^l\Lambda^m\bar{\Lambda}^n\gamma_a\Lambda_n\gamma^a\Lambda_{(i}\varepsilon_{j|l|}\varepsilon_{k)m}\;.
 \end{align}
\end{widetext}
\section{Field redefinitions and simplifying the supersymmetry transformation laws}\label{field-redef}
The supersymmetry transformation of the multiplet presented in the previous section has the advantage of the fields (with the exception of $\Lambda_{i}$) being S-invariant, which can sometimes be useful. However, the Q-supersymmetry transformations turn out to be extremely complicated and non-linear. In this section, we will present a field redefinition in which the supersymmetry transformation becomes simpler and linear with the only non-linear terms arising from the background Weyl multiplet. The field redefinitons that we find are:
 \begin{align}\label{fieldredef1}
 V &=\phi^{-2}\;,\nonumber \\
 \psi^{i}&=\phi^{-2}\Lambda^{i} \;,\nonumber \\
 K^{ij}&=\phi^{-2}\bar{E}^{ij}+\frac{1}{2}\phi^{-2}\bar{\Lambda}^{i}\Lambda^{j}\;,\nonumber \\
A_{a}{}^{i}{}_{j}&=\phi^{-2}S_{a}{}^{i}{}_{j}-\frac{1}{2}\phi^{-2}\left[\bar{\Lambda}^{i}\gamma_{a}\Lambda_{j}-\frac{1}{2}\delta^{i}_{j}\bar{\Lambda}^{m}\gamma_{a}\Lambda_{m}\right]\;,\nonumber \\
\xi_{ijk}&=\phi^{-2}\Xi_{ijk}+3\phi^{-2}\Lambda_{(i}E_{jk)}+3\phi^{-2}\slashed{S}^{m}{}_{(i}\Lambda^{n}\varepsilon_{j|m|}\varepsilon_{k)n}\nonumber \\
&\quad -\frac{3}{4}\phi^{-2}\bar{\Lambda}^{m}\gamma^{a}\Lambda_{(i}\gamma_{a}\Lambda^{n}\varepsilon_{j|m|}\varepsilon_{k)n}\;,\nonumber \\
 \mathcal{C}_{ijkl}&=\phi^{-2}C_{ijkl}+12 \phi^{-2}E_{(ij}\bar{E}_{kl)}+2\phi^{-2}\bar{\Lambda}_{(i}\Xi_{jkl)}\nonumber \\
 &\quad +3\phi^{-2}S^{n}{}_{(i}\cdot S^{m}{}_{j}\varepsilon_{k|m|}\varepsilon_{l)n}+3\phi^{-2}\bar{\Lambda}_{(i}\Lambda_{j}E_{kl)}\nonumber \\
 &\quad +3\phi^{-2}\varepsilon_{im}\varepsilon_{jn}\varepsilon_{kp}\varepsilon_{lq}\bar{\Lambda}^{(m}\Lambda^{n}E^{pq)}\nonumber \\
 & \quad +2\phi^{-2}\varepsilon_{im}\varepsilon_{jn}\varepsilon_{kp}\varepsilon_{lq}\bar{\Lambda}^{(m}\Xi^{npq)}\nonumber \\
 &\quad -6\phi^{-2}\bar{\Lambda}^{n}\slashed{S}^{p}{}_{(i}\Lambda_{j}\varepsilon_{k|n|}\varepsilon_{l)p}+\frac{3}{2}\bar{\Lambda}_{(i}\Lambda_{j}\bar{\Lambda}^{n}\Lambda^{p}\varepsilon_{k|n|}\varepsilon_{l)p}\;.
 \end{align}
 Here, we have defined:
 \begin{align}\label{Ebardef}
 \bar{E}^{ij}=\varepsilon^{ik}\varepsilon^{jl}E_{kl}\;.
 \end{align}
The above redefined fields have the same reality properties, SU(2) irreps, chiral weight and chirality as the corresponding old fields. But the Weyl weight is shifted to a negative value as shown in Table-\ref{Table-New-redef} because of the multiplication by an overall $\phi^{-2}$.
 \begin{table}\caption{}\label{Table-New-redef}
\begin{center}
  \begin{tabular}{ | p{1cm}|p{2cm}|p{1cm}|p{1cm}|p{2cm}| }
    \hline
    Field & SU(2) Irreps & Weyl weight (w) & Chiral weight (c) & Chirality (Fermions) \\ \hline
   V & $\bf{1}$ & -2 & 0 & -- \\ \hline
   $A_{a}{}^{i}{}_{j}$ & $\bf{3}$ & -1 & 0 & -- \\ \hline
   $K^{ij}$ & $\bf{3}$ & -1 & -1 & -- \\ \hline
   $\mathcal{C}_{ijkl}$ & $\bf{5}$ & 0 & 0 & -- \\ \hline
   $\Lambda_{i}$ & $\bf{2}$ & -3/2 & 1/2 & +1 \\ \hline
   $\xi_{ijk}$ & $\bf{4}$ & -1/2 & -1/2 & +1 \\ \hline
  \end{tabular}
\end{center}
\end{table}
The Q and S-supersymmetry transformation of the multiplet in terms of the redefined fields take the following simple form:
\begin{widetext}
\begin{align}\label{Q-susy_new}
 \delta V&=\bar{\epsilon}_{k}\psi^{k}+h.c.\;, \nonumber \\
 \delta\psi^{i}&=\slashed{D}V\epsilon^{i}-\slashed{A}^{i}{}_{j}\epsilon^{j}-2K^{ij}\epsilon_{j}-2V\eta^{i}\;, \nonumber\\
 \delta K^{ij}&=2V\bar{\epsilon}^{(i}\chi^{j)}-\frac{2}{3}\bar{\epsilon}^{(i}\slashed{D}\psi^{j)}+\frac{1}{3}\bar{\epsilon}^{k}\xi_{lmk}\varepsilon^{il}\varepsilon^{jm}+\frac{1}{12}\bar{\epsilon}^{(i}\gamma\cdot T^{-}\psi_{l}\varepsilon^{j)l}-2\bar{\eta}^{(i}\psi^{j)}\;,\nonumber \\
 \delta A_{a}{}^{i}{}_{j}&=V\bar{\epsilon}_{j}\gamma_{a}\chi^{i}+\frac{2}{3}\bar{\epsilon}_{j}\gamma_{a}\slashed{D}\psi^{i}-2\bar{\epsilon}_{j}D_{a}\psi^{i}-\frac{1}{3}\varepsilon^{li}\varepsilon^{nk}\bar{\epsilon}_{n}\gamma_{a}\xi_{ljk}+\frac{1}{24}\bar{\epsilon}_{j}\gamma_{a}\gamma\cdot T^{-}\psi_{k}\varepsilon^{ik}-\bar{\eta}_{j}\gamma_{a}\psi^{i}-(h.c;traceless)\;,\nonumber \\
\delta \xi_{ijk}&=\frac{3}{2}D_{a}A^{a}{}^{l}{}_{(i}\varepsilon_{j|m|}\varepsilon_{k)l}\epsilon^{m}-3D_{a}A_{b}{}^{l}{}_{(i}\varepsilon_{j|m|}\varepsilon_{k)l}\gamma^{ab}\epsilon^{m}-\frac{3}{2}V\gamma\cdot R(V)^{l}{}_{(i}\varepsilon_{j|m|}\varepsilon_{k)l}\epsilon^{m}+6\slashed{D}K^{lm}\epsilon_{(i}\varepsilon_{j|l|}\varepsilon_{k)m}-\mathcal{C}_{ijkl}\epsilon^{l}\nonumber \\
&\quad -\frac{3}{4}\gamma\cdot T^{-}\epsilon^{n}K_{(ij}\varepsilon_{k)n}-\frac{3}{2}\bar{R}(Q)_{ab}^{l}\psi_{(i}\gamma^{ab}\epsilon^{m}\varepsilon_{j|l|}\varepsilon_{k)m}-\frac{3}{2}\bar{\chi}^{l}\psi_{(i}\epsilon^{m}\varepsilon_{j|l|}\varepsilon_{k)m}+\frac{3}{2}\bar{\chi}^{l}\gamma_{ab}\psi_{(i}\gamma^{ab}\epsilon^{m}\varepsilon_{j|l|}\varepsilon_{k)m}\nonumber \\
&\quad +\frac{3}{2}\bar{\chi}_{(i}\psi^{l}\epsilon^{m}\varepsilon_{j|l|}\varepsilon_{k)m}+3 \bar{\chi}^{m}\gamma_{a}\psi^{l}\gamma^{a}\epsilon_{(i}\varepsilon_{j|l|}\varepsilon_{k)m}-6K^{nm}\eta_{(i}\varepsilon_{j|n|}\varepsilon_{k)m}-6\slashed{A}^{m}{}_{(i}\eta^{n}\varepsilon_{j|m|}\varepsilon_{k)n}\;,\nonumber \\
\delta \mathcal{C}_{ijkl}&=\bar{\epsilon}_{(i}\tilde{\Gamma}_{jkl)}+\varepsilon_{im}\varepsilon_{jn}\varepsilon_{kp}\varepsilon_{lq}\bar{\epsilon}^{(m}\tilde{\Gamma}^{npq)}-4\bar{\eta}_{(i}\xi_{jkl)}\;,
 \end{align}
 \end{widetext}
 where,
  \begin{align}\label{Gammadefnew}
  \tilde{\Gamma}_{ijk}&=-2\slashed{D}\xi_{ijk}+12\chi_{(i}K^{lm}\varepsilon_{j|l|}\varepsilon_{k)m}\;.
  \end{align}
  We can see that upon switching off the background Weyl multiplet fields, we recover the supersymmetry transformation of the multiplet presented in \cite{Howe:1982tm}, which was introduced in flat superspace as a real scalar superfield $V$ satisfying the superspace constraints:
 \begin{align}\label{superspace}
 D_{\alpha\beta}V=\left[D_{\alpha}^{i},\bar{D}_{\dot{\alpha}i}\right]V=0\;.\nonumber \\
 \end{align}
 Thus the multiplet that we have found is a generalization of the flat-space multiplet of \cite{Howe:1982tm}, to conformal supergravity background. 
 
\section{Constraining the multiplet}\label{con_mul}
In this section, we will show how to consistently constrain the real scalar multiplet to obtain a restricted multiplet with 8+8 components. For this purpose, we use the fields of Table-\ref{Table-New}. This discussion is in line with the one on N=2 chiral multiplet in the literature \cite{deRoo:1980mm}. The chiral multiplet of N=2 supergravity is a large multiplet with 16+16 components. It is known that a chiral multiplet of Weyl weight $w=1$ for the lowest component can be consistently reduced by imposing a reality constraint on an SU(2) triplet of fields $B_{ij}$ belonging to the chiral multiplet. Analogously, the 24+24 component real scalar multiplet presented in this paper can be reduced to an 8+8 matter multiplet, by imposing a reality constraint on the field $E_{ij}$. However, since $E_{ij}$ has a non trivial chiral weight $c=-1$ (as opposed to $B_{ij}$ of chiral multiplet which has $c=0$ if $w=1$ for the lowest component), we can only constrain it to a ``real'' field up to an overall phase factor, which carries the chiral weight, i.e we can constrain $E_{ij}$ to take the following form: 
\begin{align}\label{const}
E_{ij}=e^{-i\sigma/2}\mathcal{L}_{ij}\;,
\end{align}
where $\mathcal{L}_{ij}$ satisfies the reality constraint:
\begin{align}\label{const1}
\mathcal{L}^{ij}=\varepsilon^{ik}\varepsilon^{jl}\mathcal{L}_{kl}\;.
\end{align}
Trivially, under chiral transformation, the phase $\sigma$ transforms as:
\begin{align}\label{const2}
\delta_{U(1)}(\Lambda)\sigma=2\Lambda\;.
\end{align}
As a consequence of \eqref{const}, $E_{ij}$ and its complex conjugate $\bar{E}_{ij}\equiv\varepsilon_{ik}\varepsilon_{jl}E^{kl}$ are related by the constraint:
\begin{align}\label{const3}
\mathcal{R}_{ij}:=\bar{E}_{ij}-e^{i\sigma}E_{ij}=0\;.
\end{align}
The above constraint can be taken as the starting point of a procedure to obtain the full set of constraints by applying supersymmetry. For the supersymmetry transformation of $\mathcal{R}_{ij}$, we get:
\begin{align}\label{const4}
\delta\mathcal{R}_{ij}&=\bar{\epsilon}_{(i}\Theta_{j)}+\bar{\epsilon}_{k}\Upsilon^{lmk}\varepsilon_{il}\varepsilon_{jm}\nonumber \\
&\quad -e^{-i\sigma}\varepsilon_{ik}\varepsilon_{jl}\left(\bar{\epsilon}^{(k}\Theta^{l)}+\bar{\epsilon}^{m}\Upsilon_{mnp}\varepsilon^{kn}\varepsilon^{lp}\right)\;,
\end{align}
with
\begin{align}\label{const5}
\Theta_{j}&=2\chi_{j}-\frac{2}{3}\slashed{D}\Lambda_{j}+\frac{1}{12}\gamma.T^{+}\Lambda^{l}\varepsilon_{jl}+\frac{2}{3}\Lambda^{l}\bar{E}_{jl}\nonumber \\
&\quad+\frac{1}{3}\slashed{S}^{k}{}_{j}\Lambda_{k}-\frac{2}{3}\slashed{P}\Lambda_{j}+\frac{1}{12}\bar{\Lambda}^{m}\gamma_{a}\Lambda_{m}\gamma^{a}\Lambda_{j}\nonumber \\
&\quad -\frac{2i}{3}\zeta^{l}\bar{E}_{jl}\;,
\nonumber\\
\Upsilon^{lmk}&=\frac{1}{3}\Xi^{lmk}-\Lambda^{(l}E^{mk)}-i\zeta^{(l}E^{mk)}\;.
\end{align}
Here, we have defined $\zeta_{l}$ as the supersymmetry variation of the phase $\sigma$ :
\begin{align}\label{const6}
\delta\sigma=\bar{\epsilon}_{k}\zeta^{k}+\text{h.c.}\;.
\end{align}
Since $\mathcal{R}_{ij}=0$ is a supersymmetric invariant constraint, we need to set its supersymmetry variation to zero, i.e., $\Theta_i=0=\Upsilon_{ijk}$, to get the other set of constraints. From the first identity, we derive the relation between $\zeta^{k}$ and the components of the dilaton matter multiplet. Specifically:
\begin{align}\label{const7}
i\zeta^{k}=\Lambda^{k}+3E^{-2}\alpha_{j}\bar{E}^{jk}\;,
\end{align}
with
\begin{align}\label{const8}
\alpha_{j}&\equiv 2\chi_{j}-\frac{2}{3}\slashed{D}\Lambda_{j}+\frac{1}{12}\gamma.T^{+}\Lambda^{l}\varepsilon_{jl}+\frac{1}{3}\slashed{S}^{k}{}_{j}\Lambda_{k}-\frac{2}{3}\slashed{P}\Lambda_{j}\nonumber \\
&\quad +\frac{1}{12}\bar{\Lambda}^{m}\gamma_{a}\Lambda_{m}\gamma^{a}\Lambda_{j}\;,\nonumber \\
E^2 &\equiv E^{ij}E_{ij}\;.
\end{align}
The constraint $\Upsilon^{lmk}=0$, together with \eqref{const7}, gives an expression for $\Xi^{ijk}$ completely determined in terms of the lower weight fields:
\begin{align}\label{xi}
\Xi^{ijk}&=9E^{-2}\alpha_{l}E^{(ij}\bar{E}^{kl)}+6\Lambda^{(i}E^{jk)}\;.
\end{align}
At this point, one can check that $\delta_S \Xi_{ijk}=0$ by using the S-transformation of $\alpha_j$:
\begin{align*}
\delta_S \alpha_{j}=\frac{8}{3}\bar{E}_{jk}\eta^{k}\;.
\end{align*}
Also from (\ref{const7}), one finds that $\delta_S\zeta^i=-2{\rm i}\eta^i$, and hence the $\left[\delta_Q,\delta_S\right]$ commutator acts on $\sigma$ as:
\begin{align}\label{const9}
\left[\delta_S(\eta), \delta_Q(\epsilon)\right]\sigma=-2i\bar{\epsilon}_{k}\eta^{k}+2i\bar{\epsilon}^{k}\eta_{k}\;.
\end{align}
This is consistent with the U(1) transformation of $\sigma$ given in (\ref{const2}).

Now, since $\Xi_{ijk}$ is completely determined in terms of the lower weight fields, the number of independent fermionic components of the multiplet is reduced to 8 ($\Lambda_i$).

We obtain further constraints on the bosonic fields by setting the supersymmetry variation of $\Upsilon_{ijk}$ to zero. For simplicity, we present only the bosonic terms:
\begin{align}\label{susyferm}
\delta \Upsilon_{ijk}&=A_{ijkl}\epsilon^{l}+B_{(ij}\varepsilon_{k)l}\epsilon^{l}+D^{ab}{}_{ijkl}\gamma_{ab}\epsilon^{l}\nonumber \\
&\quad +F^{ab}{}_{(ij}\varepsilon_{k)l}\gamma_{ab}\epsilon^{l}+\slashed{G}_{(ij}\epsilon_{k)}+\slashed{J}_{ijkl}\epsilon_{m}\varepsilon^{lm}\;,
\end{align}
with
\begin{align}\label{const10}
A_{ijkl}&=\frac{1}{3}\left[-C_{ijkl}-12E^{-2}D_{a}P^{a}E_{(ij}\bar{E}_{kl)} \right. \nonumber \\
&\quad \left.-18E^{-2}DE_{(ij}\bar{E}_{kl)}-6E^{-2}D_{a}S^{a}{}^{m}{}_{(i}E_{jk}\bar{E}_{l)m}\right. 
\nonumber\\
&\quad \left.-12E^{-2}P_{a}P^{a}E_{(ij}\bar{E}_{kl)}+6E_{(ij}\bar{E}_{kl)}\right. \nonumber \\
&\quad-\left.12E^{-2}P_{a}S^{a}{}^{m}{}_{(i}E_{jk}\bar{E}_{l)m}\right.\nonumber \\
&\quad \left.-\frac{3}{2}E^{-2}S^{n}{}_{m}.S^{m}{}_{n}E_{(ij}\bar{E}_{kl)}\right.]+{\rm fermions}\;, \nonumber\\
B_{ij}&=\mathcal{L}^{-2}\mathcal{L}_{ij}\mathcal{L}_{kl}\mathcal{G}^{kl}+{\rm fermions}\;,\nonumber\\
F^{ab}{}_{ij}&=\mathcal{H}_{abij}-\mathcal{L}^{-2}\mathcal{L}_{ij}\mathcal{L}^{kl}\mathcal{H}_{abkl}+{\rm fermions}\;,\nonumber\\
D^{ab}{}_{ijkl}&=\frac{4}{3}\mathcal{L}^{-2}\mathcal{H}^{ab}{}_{m(i}\mathcal{L}_{jk}\mathcal{L}_{l)n}\varepsilon^{mn}+{\rm fermions}\;,\nonumber\\
J^{a}_{ijkl}&=\frac{4}{3}\mathcal{L}^{-2}G^{a}_{n(i}\mathcal{L}_{jk}\mathcal{L}_{l)m}\varepsilon^{mn}+{\rm fermions}\;,\nonumber\\
G^{a}_{ij}&=3e^{-i\sigma/2}\left(D^{a}\mathcal{L}_{ij}-\left(\mathcal{L}^{-1}D^{a}\mathcal{L}\right)\mathcal{L}_{ij}\right. \nonumber \\
&\quad \left. -S^{a}{}^{k}{}_{(i}\mathcal{L}_{j)k}\right)+{\rm fermions}\;.
\end{align}
In the above expressions, we have defined\footnote{The indices of $S_{a}{}^{i}{}_{j}$ and $R(V)_{ab}{}^{i}{}_{j}$ are raised/lowered by using the rule $A_{ij}=\varepsilon_{ik}A^{k}{}_{j}$ and $A^{ij}=A^{i}{}_{k}\varepsilon^{kj}$.}:
\begin{align}\label{const11}
\mathcal{G}^{ij}&=\frac{3}{2}\left[D_a S^{a}{}^{ij}+2P_{a}S^{a}{}^{ij}\right]\;,\nonumber\\
\mathcal{H}_{ab}{}_{ij}&=-\frac{3}{2}\left[D_{[a}S_{b]}{}_{ij}+\frac{1}{2}S_{a}{}_{m(i}S_{b}{}^{m}{}_{j)}+\frac{1}{2}R(V)_{ab}{}_{ij}\right]\;.
\end{align}
By setting \eqref{susyferm} to zero, we obtain the following set of constraints:
\begin{align}\label{const12}
A_{ijkl}&=0\;, \nonumber \\
\mathcal{G}^{ij}\mathcal{L}_{ij}&=0\;, \nonumber\\
\mathcal{H}_{ab}{}_{ij}&=\mathcal{L}^{-2}\mathcal{L}_{ij}\mathcal{L}^{mn}\mathcal{H}_{ab}{}_{mn}\;, \nonumber \\
G^{a}_{ij}&=0\;.
\end{align}
The first constraint fixes $C_{ijkl}$ in terms of the lower weight field. The third constraint in (\ref{const12}) is not an independent one, since it is equivalent to $\mathcal{L}^{lm}\varepsilon_{jm} D_{[b}{G}_{a]}{}_{il}=0$. Hence, we are left with a scalar (1) and a vector-tensor (12) constraints, which read explicitly: 
\begin{align}\label{const13}
\left(D_a S^{a}{}^{ij}+2P_{a}S^{a}{}^{ij}\right)\mathcal{L}_{ij}&=0\;,\nonumber \\
\left(D^{a}\mathcal{L}_{ij}-\left(\mathcal{L}^{-1}D^{a}\mathcal{L}\right)\mathcal{L}_{ij}-S^{a}{}^{k}{}_{(i}\mathcal{L}_{j)k}\right)&=0\;.
\end{align}
At this point, it is trivial to check that the second constraint above encompasses only 12-4=8 independent components, since a contraction with $\mathcal{L}^{ij}$ gives zero identically. Together, \eqref{const13} constitute a set of 9 independent constraints on $S_{a}{}^{i}{}_{j}$. Hence, $S_{a}{}^{i}{}_{j}$ has only 3 independent components. Furthermore, the second identity in \eqref{const13} can be solved for $S_{a}{}^{i}{}_{j}$ in terms of a 4-vector $\mathcal{H}_{a}$. To see this, consider the set of equalities:
\begin{align}\label{S}
S_{a}{}^{k}{}_{(i}\mathcal{L}_{j)k}&=S_{a}{}^{k}{}_{i}\mathcal{L}_{jk}-S_{a}{}^{k}{}_{[i}\mathcal{L}_{j]k}\nonumber \\
&=S_{a}{}^{k}{}_{i}\mathcal{L}_{jk}-\frac{1}{2}S_{a}{}^{km}\mathcal{L}_{mk}\varepsilon_{ij}
\nonumber\\
&=S_{a}{}^{k}{}_{i}\mathcal{L}_{jk}-\frac{1}{2}\phi^{-2}\mathcal{L}\mathcal{H}_a \varepsilon_{ij}\;,
\end{align}
where, we defined $S_{a}{}^{mk}\mathcal{L}_{mk}=\phi^{-2}\mathcal{L}\mathcal{H}_{a}$. The pre-factor $\phi^{-2}\mathcal{L}$ is chosen conveniently so that $\mathcal{H}_a$ has Weyl weight +3 and it satisfies a simple Bianchi identity as we will see below. 
Substituting \eqref{S} into the second identity in (\ref{const13}) and contracting with $\mathcal{L}^{lj}$, we can solve explicitly for $S_{a}{}^{i}{}_{j}$:
\begin{align}\label{S2}
S_{a}{}^{i}{}_{j}&=\mathcal{L}^{-2}\left[\mathcal{L}^{im}D_{a}\mathcal{L}_{jm}-\mathcal{L}_{jm}D_{a}\mathcal{L}^{im}\right]\nonumber \\
&\quad +\phi^{-2}\mathcal{L}^{-1}\mathcal{H}_{a}\mathcal{L}^{ik}\varepsilon_{kj}\;.
\end{align}
Plugging this solution in the first identity in (\ref{const13}), we obtain a Bianchi identity on $\mathcal{H}_a$, i.e.:
\begin{align}\label{Hconst}
D_{a}\mathcal{H}^{a}=0\;.
\end{align}
Thus, $\mathcal{H}_{a}$ can be interpreted as the dual of a 3-form field strength corresponding to a 2-form gauge field, up to possible fermionic terms.

To sum up, we have seen that, $S_{a}{}^{i}{}_{j}$ has 3 independent components which can be traded off for the dual of a three form field strength $\mathcal{H}_{a}$ satisfying a Bianchi identity. Together with the real scalar field $\phi$ (1), the phase $\sigma$ (1) and $\mathcal{L}_{ij}$ (3), they make up the total 8 bosonic components expected for a minimal matter multiplet of N=2 supergravity.

\section{The embedding of the tensor multiplet in the real scalar multiplet}\label{tensor_new}
In this section, we want to show the precise embedding of the N=2 tensor multiplet in the real scalar multiplet, which satisfies the constraint discussed in the previous section. This will show the equivalence between the tensor multiplet and the restricted real scalar multiplet. For completeness we write the full supersymmetry transformation of the tensor multiplet in a conformal supergravity background \cite{deWit:1982na,deWit:2006gn}:
\begin{align}\label{tensor_susy}
\delta L_{ij}&=2\bar{\epsilon}_{(i}\varphi_{j)}+2\varepsilon_{ik}\varepsilon_{jl}\bar{\epsilon}^{(k}\varphi^{l)} \;, \nonumber \\
\delta\varphi^{i}&=\slashed{D}L^{ij}\epsilon_{j}+\slashed{H}\varepsilon^{ij}\epsilon_{j}-G\epsilon^{i}+2L^{ij}\eta_{j}\;, \nonumber \\
\delta G&=-2\bar{\epsilon}_{i}\slashed{D}\varphi^{i}-6\bar{\epsilon}_{i}\chi_{j}L^{ij}+\frac{1}{4}\varepsilon^{ij}\bar{\epsilon}_{i}\gamma\cdot T^{+}\varphi_{j}+2\bar{\eta}_{i}\varphi^{i}\;, \nonumber \\
\delta E_{\mu\nu}&=i\bar{\epsilon}^{i}\gamma_{\mu\nu}\varphi^{j}\varepsilon_{ij}+2iL_{ij}\varepsilon^{jk}\bar{\epsilon}^{i}\gamma_{[\mu}\psi_{\nu]k}+\text{h.c}\;.
\end{align}
Here we have defined $H_{a}$ as the dual of the 3-form field strength $H_{abc}$ as shown below:
\begin{align}\label{Hdual}
H_{a}=\frac{i}{6}\varepsilon_{abcd}H^{bcd}\;,
\end{align}
where $H_{\mu\nu\rho}$ is the super-covariant field strength associated with tensor gauge field $E_{\mu\nu}$ given as:
\begin{align}\label{Hdef}
H_{\mu\nu\rho}&=3\partial_{[\mu}E_{\nu\rho]}-\frac{3i}{2}\bar{\psi}_{[\mu}{}^{i}\gamma_{\nu\rho]}\varphi^{j}\varepsilon_{ij}+\frac{3i}{2}\bar{\psi}_{[\mu}{}_{i}\gamma_{\nu\rho]}\varphi_{j}\varepsilon^{ij}\nonumber \\
&\quad -\frac{3i}{2}L_{ij}\varepsilon^{jk}\bar{\psi}_{[\mu}{}^{i}\gamma_{\nu}\psi_{\rho]k}\;.
\end{align}
Now, it is possible to verify that the following combinations of the tensor multiplet fields,
\begin{widetext}
\begin{align}\label{iden}
\phi^{4}&=L^2\;, \nonumber\\
\Lambda^{i}&=-2L^{-2}L^{ij}\varphi_{j}\;, \nonumber\\
E_{ij}&=L^{-4}L_{ij}L^{kl}\bar{\varphi}_{k}\varphi_{l}-L^{-2}\bar{G}L_{ij}\;, \nonumber\\
S_{a}{}^{i}{}_{j}&=2L^{-2}H_{a}L^{ik}\varepsilon_{kj}+4L^{-4}L^{ik}L_{jm}\bar{\varphi}^{m}\gamma_{a}\varphi_{k}-L^{-2}\bar{\varphi}^{i}\gamma_{a}\varphi_{j}-\frac{1}{2}L^{-2}\delta^{i}_{j}\bar{\varphi}^{m}\gamma_{a}\varphi_{m}+L^{-2}\left(L^{ik}D_{a}L_{jk}-L_{jk}D_{a}L^{ik}\right)\;, \nonumber\\
\Xi_{ijk}&= -24L^{-6} L^{mn}\bar{\varphi}_m\varphi_nL_{(ij}L_{k)l}\varphi^l+6L^{-4}\bar{\varphi}_l\varphi_{(i}L_{jk)}\varphi^l-6L^{-4}L^{lm}\slashed{D}L_{l(i}L_{jk)}\varphi_m +6L^{-4}L^{mn}\slashed{H}\varphi_nL_{(ij}\varepsilon_{k)m}\nonumber\\
&\quad +12L^{-4}\bar{G}L_{(ij}L_{k)l}\varphi^l+6L^{-2}\slashed{D}\varphi_{(i}L_{jk)}+18L^{-2}L_{(ij}L_{k)l}\chi^l-\frac{3}{4}L^{-2}\gamma\cdot T^{-}\varphi^lL_{(ij}\varepsilon_{k)l}\;,\nonumber \\
C_{ijkl}&=-18L^{-2}DL_{(ij}L_{kl)}+6L^{-4}G\bar{G}L_{(ij}L_{kl)}+6L^{-4}H^aH_aL_{(ij}L_{kl)}-12L^{-4}H^aD_aL^{mn}\varepsilon_{m(i}L_{jk}L_{l)n}\nonumber\\
&\quad-6L^{-2}D_aD^aL_{(ij}L_{kl)}+6L^{-4}L^{mn}D_aL_{mn}D^aL_{(ij}L_{kl)}-3L^{-2}D^aL^{mn}D_aL_{mn}L_{(ij}L_{kl)}\nonumber\\
&\quad-9L^{-2}\bar{\chi}^m\varphi^n L_{(ij}\varepsilon_{k|m|}\varepsilon_{l)n}-36L^{-4}\bar{\chi}^m\varphi^nL_{(ij}L_{k|m|}L_{l)n}-36L^{-4}L^{mn}\bar{\chi}_m\varphi_nL_{(ij}L_{kl)}+9L^{-2}\bar{\chi}_{(i}\varphi_jL_{kl)}\nonumber\\
&\quad+6L^{-4}G\bar{\varphi}_{(i}\varphi_jL_{kl)}-18L^{-6}GL^{mn}\bar{\varphi}_m\varphi_nL_{(ij}L_{kl)} +6L^{-6}\bar{G}L_{mn}\bar{\varphi}^m\varphi^nL_{(ij}L_{kl)}-6L^{-4}\bar{G}\bar{\varphi}^m\varphi^nL_{(ij}\varepsilon_{k|m|}\varepsilon_{l)n}\nonumber\\
&\quad-24L^{-6}\bar{G}\bar{\varphi}^m\varphi^nL_{(ij}L_{k|m|}L_{l)n}-6L^{-4}\bar{\varphi}^m\gamma^a\varphi_{(i}L_{jk}D_aL_{l)m}
+36L^{-6}\bar{\varphi}^m\gamma^a\varphi_nL^{ns}D_aL_{s(i}L_{jk}L_{l)m}\nonumber\\
&\quad -6L^{-4}\bar{\varphi}^m\gamma^a\varphi_mD_aL_{(ij}L_{kl)}+6L^{-4}\bar{\varphi}^m\slashed{H}\varphi_{(i}L_{jk}\varepsilon_{l)m}-36L^{-6}\bar{\varphi}^m\slashed{H}\varphi_nL^{ns}L_{(ij}L_{k|m|}\varepsilon_{l)s}\nonumber\\
&\quad-12L^{-4}L^{st}\varphi_s\slashed{D}\varphi^mL_{(ij}\varepsilon_{k|t|}\varepsilon_{l)m}-12L^{-4}\bar{\varphi}^m\slashed{D}\varphi_{(i}L_{jk}L_{l)m}+\frac{3}{4}L^{-4}\varepsilon_{mn}\bar{\varphi}^m\gamma\cdot T^-\varphi^nL_{(ij}L_{kl)}\nonumber\\
&\quad+\frac{3}{4}L^{-4}\varepsilon^{mn}\bar{\varphi}_m\gamma\cdot T^+\varphi_nL_{(ij}L_{kl)}-36L^{-6}\bar{\varphi}^m\varphi^n\bar{\varphi}_m\varphi_nL_{(ij}L_{kl)}-36L^{-6}L_{mn}\bar{\varphi}^m\varphi^n\bar{\varphi}_{(i}\varphi_jL_{kl)}\nonumber\\
&\quad+36L^{-6}\bar{\varphi}^m\varphi^n\bar{\varphi}_m\varphi_{(i}L_{jk}L_{l)n}+90L^{-8}L^{mn}\bar{\varphi}_m\varphi_nL_{st}\bar{\varphi}^s\varphi^tL_{(ij}L_{kl)}\;,
\end{align}
\end{widetext}
transform as the components of the real scalar multiplet (\ref{Susy-transf}). One can check that the above combinations satisfy the constraints presented in the previous section. One can also check that the combinations appearing as $S_{a}{}^{i}{}_{j}$ and $C_{ijkl}$ above also satisfy the reality constraints \eqref{Sreality}
The relation between the tensor multiplet and the independent components of the restricted real scalar multiplet can be obtained from the identification (\ref{iden}) and are given as:
\begin{align}\label{iden1}
\phi^{4}&=L^2\;, \nonumber\\
\Lambda^{i}&=-2L^{-2}L^{ij}\varphi_{j}\;, \nonumber\\
e^{-i\sigma/2}&=\left(\frac{\mathcal{Z}}{\bar{\mathcal{Z}}}\right)^{1/2}\;, \nonumber \\
\mathcal{L}_{ij}&=\left|\mathcal{Z}\right|L_{ij}\;, \nonumber \\
\mathcal{H}_{a}&= H_{a}+\text{fermions}\;, 
\end{align}
Here $\mathcal{Z}$ is defined as the following combination of tensor multiplet fields:
\begin{align}\label{Zdef}
\mathcal{Z}&=L^{-4}L^{kl}\bar{\varphi}_{k}\varphi_{l}-L^{-2}\bar{G}\;.
\end{align}
Thus, we see that the off-shell 8+8 N=2 tensor multiplet can be embedded within the 24+24 real scalar multiplet that satisfies the constraints (\ref{const3}, \ref{xi}, \ref{const12}), mentioned in the previous section \footnote{We thank Daniel Butter for pointing out this possibility.}.
\section{Conclusions and future directions}
Superconformal couplings of the matter multiplets and the gauge multiplet are crucial in our understanding of the structure of supergravity that can arise from them. Apart from the structure of the higher derivative terms that one can obtain, they also tell us about the geometric structure of the matter couplings. For example, it is well-known that the scalar manifold corresponding to vector multiplet coupled to N=2 conformal supergravity in four dimensions, upon gauge fixing and eliminating the auxiliary fields, corresponds to a special K\"ahler manifold (see \cite{deWit:1984wbb,deWit:1984rvr,Cremmer:1984hj} and \cite{Mohaupt:2000mj} for a review). Similar geometric structure exists for other matter couplings in four-dimensional as well as five-dimensional supergravity. Hence, the discovery of new matter multiplets opens up the possibilities of gaining new insights into supergravity structures arising from them. 

In this paper, we have presented one such multiplet, the real scalar multiplet, which contains 24+24 off-shell degrees of freedom and generalizes the flat space result of \cite{Howe:1982tm} to a conformal supergravity background\footnote{Such multiplets have also recently been constructed in six dimensions \cite{Kuzenko:2017jdy}.}. We have also shown that one can impose a consistent set of constraints to reduce the multiplet to 8+8 components and as an example, we have shown the embedding of the tensor multiplet within the real scalar multiplet that satisfies these constraints. 

As a potential application of this result, one can construct a density formula in terms of the real scalar multiplet, which by virtue of the tensor multiplet embedding \eqref{iden}, will allow us to look for new couplings of tensor multiplet to conformal supergravity. This would generalize the improved tensor multiplet action obtained in \cite{deWit:1982na} using an invariant density formula that involved the chiral multiplet. We leave this study for a future work. It would also be interesting to see if results analogous to the ones presented in section \ref{tensor_new} can be obtained for vector and Weyl multiplets, and the possibility of new formulations of conformal and Poincar\'e supergravity.

A complete knowledge of the structure of supergravity arising from various matter and gauge couplings, in an off-shell formulation, is crucial to our understanding of black holes, in particular, the effective quantum corrections to their entropy. In fact, using the standard Weyl multiplet and other matter multiplets known so far, several higher derivative invariants have been constructed in four spacetime dimensions \cite{deWit:1996gjy,deWit:1996ag,deWit:2010za,Butter:2013lta} and their effect on black hole entropy have been studied \cite{deWit:2010za,Sahoo:2006rp,Banerjee:2016qvj,Butter:2014iwa}. It has also been shown that the above results are able to correctly reproduce the microscopic entropy of BPS black holes \cite{deWit:2010za,Sahoo:2006rp,Butter:2014iwa} as well as some class of non-BPS black holes \cite{Banerjee:2016qvj}. However, the result presented in this paper practically allows for the existence of new, unknown couplings in supergravity, and their connection to higher dimensional theories and string theory in general. As an inevitable consequence, the need will arise to revisit and put under scrutiny every result mentioned above (and more), based on the analysis of invariant couplings of supergravity.

\begin{acknowledgments}
 The authors would like to thank Daniel Butter and Bernard de Wit for their helpful comments on this work. We would like to thank Daniel Butter and Sergei Kuzenko for bringing the references \cite{Howe:1982tm,Stelle:1981gi,Butter:2010sc,Kuzenko:2017jdy} to our notice. We would also like to thank the organizers of the Indian National Strings Meeting (NSM) 2015 (IISER Mohali) for giving us a platform for illuminating discussions which led to the idea behind this work.
\end{acknowledgments}

\end{document}